\documentclass[fleqn,usenatbib]{mnras}

\usepackage{newtxtext,newtxmath}

\usepackage[T1]{fontenc}

\DeclareRobustCommand{\VAN}[3]{#2}
\let\VANthebibliography\thebibliography
\def\thebibliography{\DeclareRobustCommand{\VAN}[3]{##3}\VANthebibliography}

\usepackage{graphicx}	
\usepackage{amsmath}	
\usepackage{multicol}
\usepackage{deluxetable}
\usepackage{comment}

\title[X-ray Variability in Haro 11]{Resolving the Ultra-Luminous X-ray Sources in the Ly$\alpha$ Emitting Galaxy Haro 11}

\author[A. C. Gross et al.]{
Arran C. Gross,$^{1}$\thanks{E-mail: acgross@uiowa.edu}
Andrea Prestwich,$^{2}$
Philip Kaaret$^{1}$
\\
$^{1}$Department of Physics and Astronomy, University of Iowa, Van Allen Hall, Iowa City, IA 52242, USA\\
$^{2}$Harvard-Smithsonian Center for Astrophysics, 60 Garden Street, Cambridge, MA 02138, USA\\
}

\date{Accepted XXX. Received YYY; in original form ZZZ}

\pubyear{2020}

\begin{document}
\label{firstpage}
\pagerange{\pageref{firstpage}--\pageref{lastpage}}
\maketitle

\begin{abstract}

Lyman continuum and line emission are thought to be important agents in the reionization of the early universe. Haro 11 is a rare example of a local galaxy in which Ly$\alpha$ and continuum emission have escaped without being absorbed or scattered by ambient gas and dust, potentially as a consequence of feedback from its X-ray sources. We build on our previous {\it Chandra} analysis of Haro 11 by analyzing three new observations. Our subpixel spatial analysis reveals that the two previously known X-ray sources are each better modelled as ensembles of at least 2 unresolved point sources. The spatial variability of these components reveals X1 as a dynamical system where one luminous X-ray source ($L_{\rm X} \sim 10^{41}$ erg s$^{-1}$) fades as a secondary source begins to flare. These might be intermediate mass black holes or low luminosity active galactic nuclei near the center of the galaxy in the process of merging. Optical emission line diagnostics drawn from the literature suggest that while the galaxy as a whole is consistent with starburst signatures of ionization, the individual regions wherein the X-ray sources reside are more consistent with AGN/composite classification. The sources in X2 exhibit some degree of flux variability. X2a dominates the flux of this region during most observations ($L_{\rm X} \sim 6\ \times\  10^{40}$ erg s$^{-1}$), and gives the only evidence in the galaxy of a soft Ultra-Luminous X-ray source capable of high energy winds, which we suggest are responsible for allowing the coincident Ly$\alpha$ emission to escape. 

\end{abstract}




\section{Introduction}

Ultraviolet light from the earliest galaxies is thought to be responsible for transforming the intergalactic medium of the early universe from cold and neutral to warm an ionized \citep{Mesinger13}. This "Epoch of Reionization" is thought to have occurred between about 400 and 900 million years after the Big Bang \citep{Dunkley09, Bosman18}, during which time the universe contained an abundance of lower metallicity dwarf galaxies undergoing intense bouts of intense star-formation \citep{BasuZych13, Alavi16}. A profusion of UV photons in the form of Lyman continuum and line emission emanates from the young massive stars in these early galaxies \citep{Steidel99, Shapley06, Nilsson07, Mallery12}, and is thought to be the most likely cause of the reionization \citep{Loeb10, Heckman11}. Above the Lyman limit at rest-frame wavelength of 912\AA, the energy to ionize neutral Hydrogen, higher energy photons form a continuum of emission (Lyman continuum, LyC). As a Hydrogen recombination line, Ly$\alpha$ line emission (rest-frame $\lambda\sim 1215.7$\AA) is an important tracer of the intense starbursts or active galactic nuclei (AGN) needed to ionize the intergalactic medium (for a recent review, see \citet{Ouchi20}). But directly linking the emission of these high energy photons to individual sources is not possible for high-redshift early galaxies due to limits of spatial resolution.

As a means of studying Ly$\alpha$ and continuum emission in galaxies similar to those during reionization, \citet{Heckman05} and \citet{Hoopes07} constructed a sample of local proxies. Using $GALEX$ and the Sloan Digital Sky Survey, the sample of so-called Lyman Break Analogues (LBAs) consists of compact star-forming galaxies with high UV surface-brightness and low metallicity, similar to early Lyman emitting galaxies. However, determining a clear-cut relationship between the Lyman emissions and their original sources is hampered even in spatially resolved studies of local starbursts by mechanisms within the galaxies that either absorb or redistribute the UV photons. Absorption of LyC and Ly$\alpha$ photons by dust reduces the amount of flux that manages to escape the galaxy to values well below emission estimates predicted by simple models of \ion{H}{ii} regions \citep{Kunth03, MasHesse03, Ostlin09}. Further complicating the picture, Ly$\alpha$ photons are resonantly scattered by neutral hydrogen outside the ionization region of the emitting starburst \citep{Neufeld91, Hayes10, Verhamme12}, leading to extended diffusion of Ly$\alpha$ far removed from H$\alpha$ and other optical signatures \citep{Ostlin15}.

\begin{figure}
	\includegraphics[width=\columnwidth]{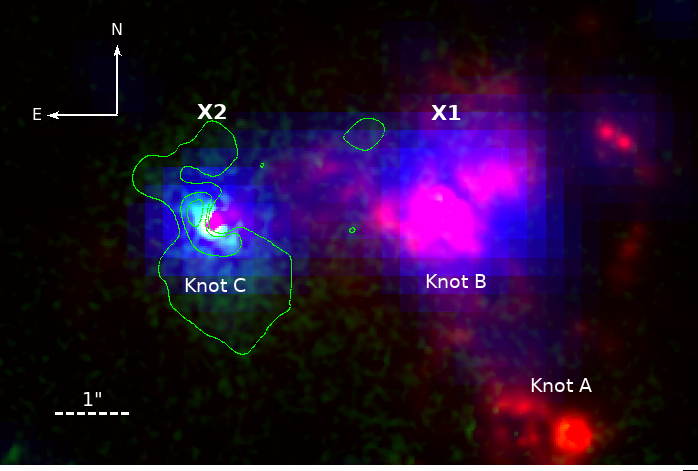}
    \caption{Haro 11 multi-band composite of H$\alpha$ emission (red), Ly$\alpha$ line emission (green) \citep{Ostlin09}, and hard band X-rays (blue, 3$-$5 keV). Both X1 and X2 appear to be coincident with star-forming regions (Knots B and C, respectively). X2 is centered on a bright region of concentrated Ly$\alpha$ emission, which exhibits a bi-conical outflow shown by the green contours. The widest contour highlights the diffuse low surface brightness Ly$\alpha$ halo, suggested by \citet{Ostlin09} to be a direct consequence of resonance scattering. The X-ray image is a stacked image of the 4 observations, at 1/2 native pixel size. All images have been smoothed with a Gaussian kernel, and are arbitrarily scaled to highlight overlap. }
    \label{fig:Lya}
\end{figure}

The means by which the Lyman photons escape a galaxy remains an outstanding question central to our understanding of both the early and present universe; however, it is generally agreed that some form of mechanical feedback is required to blow away and disperse the absorbing media to create channels through which the photons can escape unabated \citep{Orsi12, Wofford13}. Feedback from supernovae ejecta and winds generated by the starburst can create regions of partial ionization ("super-bubbles") \citep{TenorioTagle99, Hayes10, Heckman11}, resulting in localized escape routes dubbed the "picket-fence" model \citep{Keenan17}. Accretion onto compact objects such as black holes can also be a source of mechanical feedback, via jets and accretion-induced outflows, which generates power comparable to the radiative luminosity \citep{Gallo05, Justham12}. Observations of star-forming galaxies provide evidence that this power can sometimes exceed the luminosity \citep{Kaaret17b}, even by up to factors of 10$^{4}$ of the X-ray luminosity \citep{Pakull10}. Such accreting sources might be an X-ray binary (XRB; see \citet{Remillard06} for a review) in which a stellar mass black hole is fed by a massive companion star, or perhaps a central supermassive black hole (SMBH) triggered to accrete as an active galactic nucleus (AGN). Between these extremes might exist intermediate mass black holes (IMBHs) which might accrete as ultraluminous X-ray sources (ULXs; defined in \citet{Makishima00}, but see \citet{Kaaret17b} and for a review) or low luminosity AGN (LLAGN) \citep{Chilingarian18}, but whose existence has hitherto not been definitively confirmed (see \citet{Abbott20a, Abbott20b} for a recent potential detection of an IMBH (142 $M_{\sun}$) via gravitational waves). Such IMBHs found in young galaxies, particularly those in the midst of a galactic merger, might potentially undergo rapid accretion to evolve into a SMBH powering a more luminous AGN, such as a quasar capable of galaxy-wide influence. While smaller in scale, ULXs have also been shown to have significant impact on their surrounding environments via feedback from winds and jets capable of shock-exciting gas beyond levels due to supernovae and stellar winds \citep{Abolmasov07, Abolmasov11}.

Haro 11 is a well-studied blue compact dwarf (BCD) starburst galaxy in the local universe that is particularly well suited to address the topics above (see \citet{Adamo10, Hayes07, Ostlin09} and references therein). As part of the \citet{Hoopes07} sample, it has UV and physical characteristics of a LBA, and should be a good low-$z$ counterpart to Lyman $\alpha$ emitters (LAEs) at $z \gtrsim 2$ \citep{Ouchi20}. Identified directly as a local LAE \citep{Kunth03}, it is also one of only three known LyC emitters (dubbed "Lyman Continuum Leakers," LCLs) within 1000 Mpc \citep{Bergvall06, Leitet11} (the other two being Tololo 1247-232 \citep{Leitet13}, and Mrk 54 \citep{Leitherer16}). The irregularly shaped galaxy (perhaps due to an ongoing merger \citep{Ostlin09}) is host to a burst of unusually efficient star-formation \citep{Pardy16}, which is resolved into three discreet regions dubbed star-forming Knots A, B, and C by \citet{Vader93}. \citet{DeRossi18} find that the far infrared spectral energy distribution of Haro 11 is a good match to models of Population II galaxies, making it a relevant local proxy for studying processes in the early universe.

In our previous study \citep{Prestwich15}, we showed that the X-ray emission is concentrated to two unresolved sources: CXOU J003652.4-333316.95 (Haro 11 X1) coincides with Knot B, and CXOU J003652.7-333316.95 (Haro 11 X2) coincides with both Knot C and the bulk of the Ly$\alpha$ emission (see Figure \ref{fig:Lya}). X1 was shown to contain a bright ($L_{\rm 0.3-8 keV} \sim 1 \times 10^{41}$ erg s$^{-1}$) hard X-ray source ($\Gamma$ = 1.2$\pm0.2$) surrounded by a diffuse extended thermal component likely due to the surrounding starburst. We suggested that X1 could be a single accretion source, perhaps an AGN or an IMBH accreting in the low state. Conversely, X2 had a markedly softer spectrum ($\Gamma$ = 2.2$\pm0.4$) and was slightly fainter ($L_{\rm 0.3-8 keV} \sim 5 \times 10^{40}$ erg s$^{-1}$) and more spatially compact, suggestive of a soft ULX capable of super-Eddington accretion. Both sources showed negligible absorbing column densities ($N_{\rm H} \lesssim 2 \times 10^{21}$cm$^{-2}$), but did exhibit clear spectral turnovers in the fitted power laws around 3 keV, suggestive of XRB-type sources \citep{Gladstone09}.

\begin{figure*}
	\includegraphics[width=1\textwidth]{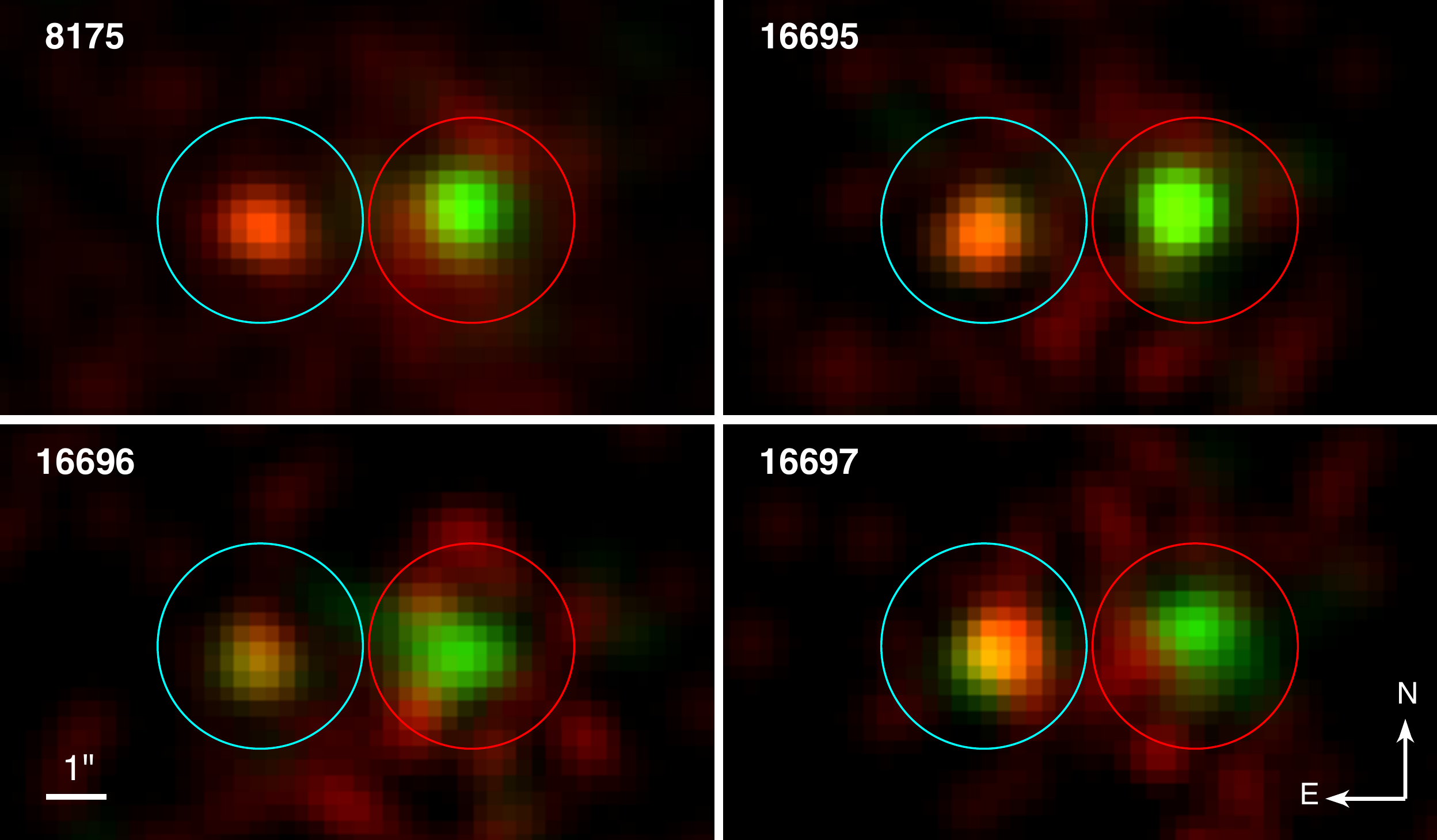}
	\centering
    \caption{Montage of smoothed, exposure-corrected false-colour images of Haro 11 for each observation. Soft X-rays (red, 0.3$-$1 keV) and hard X-rays (green, 3$-$5 keV) are scaled the same across all images. Extraction apertures are superimposed. X2 (blue circle) appears to emit primarily soft X-rays in ObsId 8175, but becomes progressive harder over the subsequent observations. X1 (red circle) appears to be dominated by hard X-rays over all observations, although its spatial extent appears to elongate during ObsId 16696. The pixels are scaled to 1/2 native size (scalebar given in bottom left corner, 1\arcsec\ $\sim$ 400 pc).  }
    \label{fig:color_smooth}
\end{figure*}

In this follow-up study to \citet{Prestwich15}, we add three additional $Chandra$ observations to the original to address the following questions about Haro 11: 1) are there multiple X-ray sources detected in either the X1 or X2 regions; and 2) what is the nature of the X-ray emission ({\it e.g.}, spectral hardness, variability over time) of those sources? We build on our previous conclusions, and aim to better constrain the physical picture of sources in Haro 11 in an effort to explain the Ly$\alpha$ and LyC emissions. In \S\ref{sec:obs} we detail our suite of observations and the data reduction. In \S\ref{sec:spectral} we analyze the regions of X1 and X2 via spectral fitting, obtaining lightcurves over the course of the 4 observations. We then conduct 2D modelling of both regions in \S\ref{sec:2D} to determine whether either region contains multiple point sources. In \S\ref{sec:discussion} we compare the number of sources found in  Haro 11 to similar nearby galaxies, discuss the optical signatures of Haro 11 X1 and X2, and synthesize possible interpretations for the physical nature of its diverse X-ray sources and their relation to the observed Ly$\alpha$ emission. We summarize our conclusions in \S\ref{sec:conclusions}. Throughout, we assume Haro 11 is at a distance of 84.0 Mpc, for an angular scale of 407 pc arcsec$^{-1}$ (NASA Extragalactic Database).

\section{X-ray Observations}
\label{sec:obs}

\begin{deluxetable}{lcc}
\tablewidth{162.80998pt}
\tablecaption{Chandra Observations}
\tablehead{ 
\colhead{ObsID} & \colhead{UT date} & \colhead{Exp. Time (ks)}}

\startdata  

08175 & 2006 Oct 28 & 54.00 \\[0.1cm]
16695 & 2015 Nov 29 & 24.74\\[0.1cm]
16696 & 2016 Sep 12 & 24.74\\[0.1cm]
16697 & 2017 Nov 24 & 23.75\\[0.1cm]

\enddata
\label{tab:obs}
\end{deluxetable}

Haro 11 was initially observed using \textit{Chandra} under ObsID 8175 in 2006, as reported in \citet{Prestwich15}. To better assess the nature and possible variability of the multiple X-ray sources in this galaxy, three subsequent observations were taken between 2015 and 2017. The observational information is given in Table \ref{tab:obs}. All observations were carried out using \textit{Chandra's} ACIS-S array, with the target position located near the aimpoint on the S3 chip in all cases.

The data were processed using CIAO version 4.12 and CALDB version 4.9.2.1 using \texttt{chandra\_repro}. The images were then aligned to ensure consistency of extraction regions. Using exposure-corrected broadband images created using the \texttt{flux\_image} task, a catalog of X-ray sources was generated for each observation using the Mexican hat wavelet algorithm, \texttt{wavdetect}. Since there did appear to be differences in morphology of Haro 11 across the observations and in multiple energy bands, we excluded Haro 11 as a whole when generating the catalog. This is necessary because of noticeable differences in both the number and location of sources found in the region by \texttt{wavdetect}. The detection catalog positions of the three subsequent observations were matched against those of the first observation since it has roughly double the exposure time of the later observations. Offsets were then applied to the evt2 files using the  \texttt{wcsmatch} and  \texttt{{}wcsupdate} tasks.

In Figure \ref{fig:color_smooth} we show a false-colour montage of the four observations. We choose two narrow energy bands to highlight the diffuse soft (0.3$-$1 keV) X-rays that permeate both sources contrasted against the spatially concentrated hard (3$-$5 keV) X-ray cores of both sources. At a glance, it is apparent that X1 is a predominantly hard source, though the exact spatial distribution of the counts does fluctuate slightly, hinting perhaps at multiple unresolved components. On the other hand, X2 is more spatially compact but shows signs of spectral hardening over the course of the observations.

\begin{figure}
	\includegraphics[width=\columnwidth]{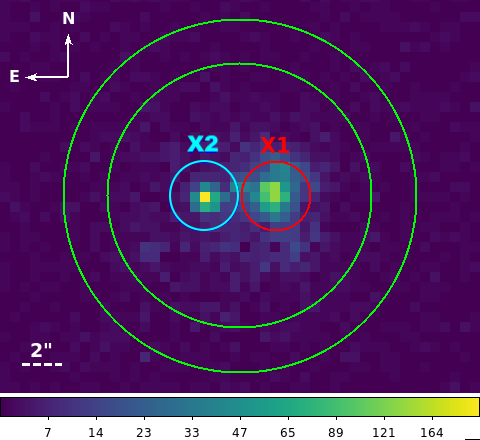}
    \caption{Stacked image of Haro 11 in broadband (0.3$-$8 keV). The extraction apertures used for all observations are superimposed. Both X1 (red) and X2 (blue) are extracted using apertures of 1.7\arcsec\ radius, and a common annular background region (green) with a width of 2.2\arcsec. The pixels are shown at native size (scalebar given in bottom left corner, 2\arcsec\ $\sim$ 800 pc). The colourbar is in units of total counts.  }
    \label{fig:stack_image}
\end{figure}

\section{Spectral Analysis}
    \label{sec:spectral}

In order to test the possibility of spectral hardening in the sources, or fluctuations in luminosity, we extract spectra for X1 and X2 for each observation. 

In Figure~\ref{fig:stack_image}, we show the stacked broadband image of Haro 11 with the apertures used for spectral extraction superimposed. For both X1 and X2, a circular aperture of radius 1.7\arcsec\ is used, which is the maximum size to avoid overlap of the targets. A common background is extracted from an annular region just outside the spatially extended diffuse emission of Haro 11, with inner and outer radii 6.5\arcsec\ and 8.7\arcsec\ out from the coordinate center, respectively. This arrangement of extraction regions is applied to all 4 observations using the  \texttt{spec\_extract} task, across the entire 0.3$-$10 keV range. We note that our choice of apertures differs from the smaller apertures used for 8175 in \citet{Prestwich15}, but for the shorter exposure images, this is partially motivated to capture as many counts as possible for binned fitting.

Because the data in many of the extraction regions are low-count regime ($\sim$200 cts.), we conduct model fitting in \texttt{Sherpa} using Cash statistics \citep{Cash79}, along with the Nelder-Mead simplex optimization method. Because of this, we do not subtract the background spectrum, but instead model it simultaneously. The fit is restricted to the 0.5$-$8 keV range, and data are binned such that there are at least 5 counts per bin. We fit the data using a power law with photo-electric absorption using Wisconsin cross-sections (\texttt{xswabs * powlaw1d}) \citep{Morrison83}. The only restriction imposed during the fitting is that the neutral Hydrogen column density, $N_{\rm H}$, cannot drop below the galactic value in the direction of Haro 11 (1.88 $\times 10^{20}$ cm$^{-2}$ \citep{Dickey90}). The only exception to this is the fit done on X1 in observation 16697, where we fix $N_{\rm H}$ to the galactic value so that the resulting covariance matrix for the fit is positive definite (necessary for flux estimate). We simultaneously model the background spectrum with a power law. We give the plots for the individual spectral fits in Appendix \ref{fig:spec}. 

\begin{figure*}
	\includegraphics[width=1\textwidth]{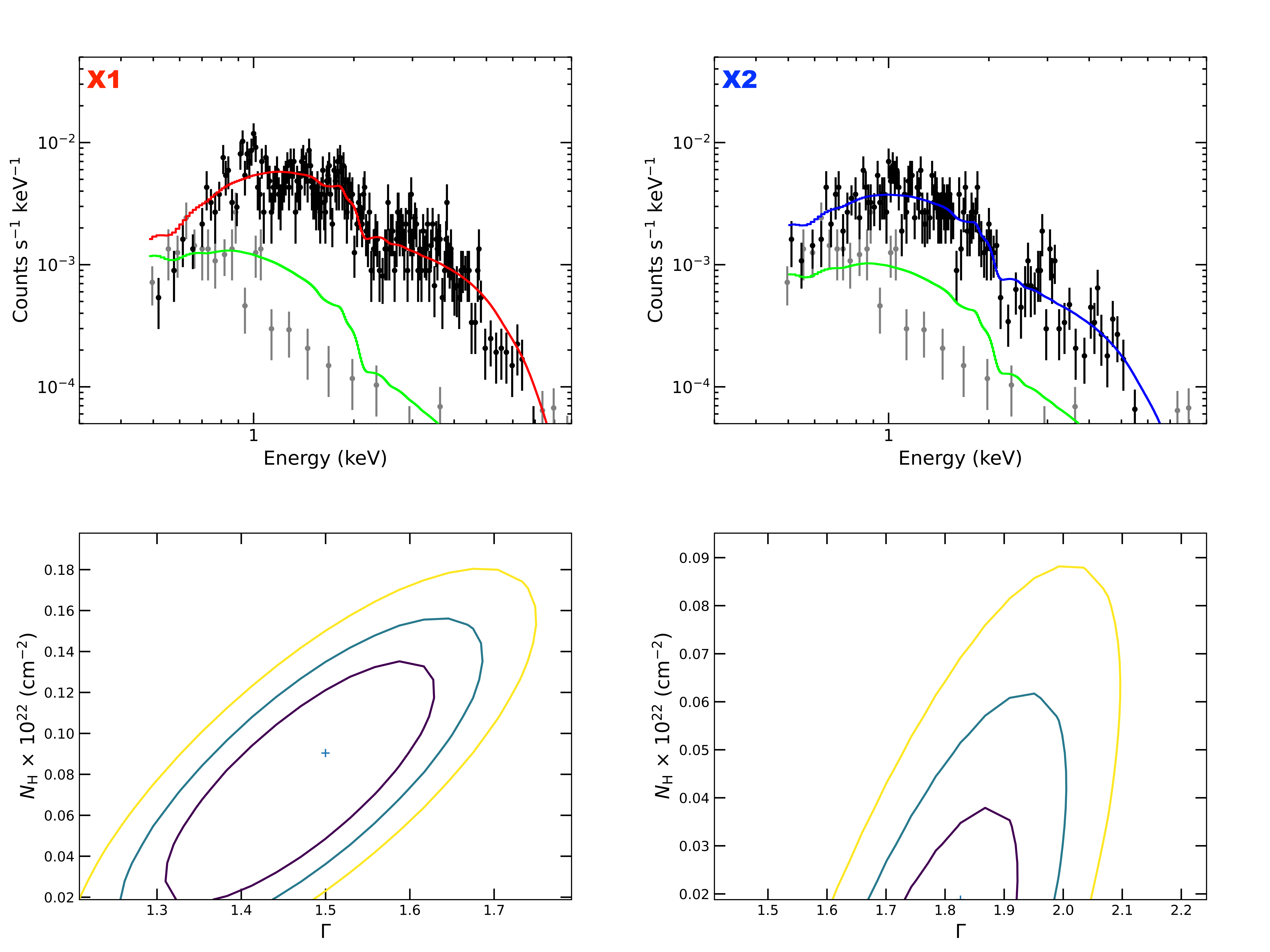}
	\centering
    \caption{{\bf Top:} Best-fit stacked X-ray spectra of Haro 11. The left panel shows the stacked spectrum for X1, and the right panel shows the stacked spectrum for X2. The stacked fits represent a median of the individual observations. The data (black points) are binned to a minimum of 5 cts/bin. The background (gray points) is not subtracted, but instead modelled simultaneously (green curve) since Cash statistics are employed. The best fit model (red curve for X1 and blue curve for X2) is an absorbed power law, with best fit parameters and 90\% confidence intervals given in Table \ref{tab:speclums}. {\bf Bottom:} Error contours for spectral fits. The contours show the behaviour and relation of the uncertainty estimates for the fitting parameters photon index ($\Gamma$) and neutral Hydrogen column density ($N_{\rm H}$). The $+$ shows the location of the best fit parameters. The uncertainty contours outward from the middle are 1$\sigma$  (purple), 2$\sigma$ (teal), and 3$\sigma$ (yellow). The 90\% confidence intervals given in Table \ref{tab:speclums} ($\sim1.6\sigma$) fall between the central two contours. The best fit for X2 is at the imposed minimum threshold for $N_{\rm H}$, hence the lower bound is unconstrained. }
    \label{fig:stacked_spec}
\end{figure*}

As a median, we also fit the stacked spectra for X1 and X2. The extracted spectra for each region are combined using the \texttt{combine\_spectra} task. These spectra are then fit using the same procedure as above. The resulting best-fits are plotted for the stacked spectra of X1 and X2 in Figure \ref{fig:stacked_spec}. Because the photon index and column density are degenerate parameters, we also show the corresponding error contours relating the uncertainty estimates of $\Gamma$ and $N_{\rm H}$. For X1, the uncertainties of these parameters are not affected by imposing a minimum threshold for $N_{\rm H}$ at the galactic value, and the best-fit values lie well above the threshold. This is not the case for X2, whose best-fit column density is at the imposed minimum for all fits. This can be seen in the error contours where the uncertainties are unbounded below the galactic value. If we remove the constraint that $N_{\rm H}$ must be at least the galactic value, the best-fit $\Gamma$ is still consistent with the constrained value for all fits; however, the estimates for $N_{\rm H}$ in X2 tend to zero, and the error contours become unbounded due to the low counts of the spectrum and choice of binning scheme. We therefore choose to keep the imposed minimum $N_{\rm H}$. These results are not surprising, as the region of Knot B has been shown to be more obscured by dust than Knot C \citep{Adamo10}. 

\begin{deluxetable}{lcccc}
\centering
\tablewidth{243.10081pt}
\tablecaption{Spectral Fitting Results}
\tablehead{ 
\colhead{ObsID} & \colhead{src cts} & \colhead{total cts}&  \colhead{$\Gamma$} & \colhead{$N_{\rm H}$}\\
\colhead{} & \colhead{0.3$-$8 keV} & \colhead{0.3$-$8 keV} &\colhead{} & \colhead{$\times$ 10$^{20}$ cm$^{-2}$}
}
\startdata 
&&X1&& \\
\hline
\vspace{0.25cm}
8175  &  686 & 773 & 1.5$^{+0.2} _{-0.2}$  & 9.0$^{+5.7} _{-5.5}$   \\\vspace{0.15cm}
16695  & 265 & 286 & 1.7$^{+0.3} _{-0.3}$  & 21.9$^{+16.1} _{-13.3}$   \\\vspace{0.15cm}
16696  &    225 & 242 & 1.4$^{+0.3} _{-0.2}$  & 4.7$^{+13.0} _{-...}$     \\\vspace{0.15cm}
16697  &  179 & 199 & 1.5$^{+0.0} _{-0.0}$  & 1.9    \\\vspace{0.15cm}
Stacked &  1355 & 1500 & 1.5$^{+0.1} _{-0.1}$  & 9.0$^{+4.4} _{-4.3}$   \\
\hline 
\hline
&&X2&& \\
\hline
\vspace{0.25cm} 
8175  &   339 & 426 & 2.0$^{+0.2} _{-0.2}$  & 1.9$^{+5.8} _{-...}$   \\\vspace{0.15cm}
16695  &   149 & 170 & 1.8$^{+0.2} _{-0.2}$  & 1.9$^{+6.0} _{-...}$    \\\vspace{0.15cm}
16696  &   101 & 118 & 1.5$^{+0.3} _{-0.3}$  & 1.9$^{+6.4} _{-...}$      \\\vspace{0.15cm}
16697  &   158 & 178 & 2.1$^{+0.3} _{-0.2}$  & 1.9$^{+6.4} _{-...}$     \\\vspace{0.15cm}
Stacked & 747 & 892 & 1.8$^{+0.1} _{-0.1}$  & 1.9$^{+2.1} _{-...}$   \\
\enddata
\label{tab:spec}
\tablecomments{Best-fit parameters of spectral fitting using an absorbed power law model. The fitting results for X1 and X2 are separated by the horizontal middle line. The top half gives the fit results of X1, and the bottom half gives the results for X2. The best-fit stacked spectra are plotted in Figure \ref{fig:stacked_spec}. The individual fitted spectra are given in the Appendix Figure \ref{fig:spec}. Error bars give the 90\% confidence interval ($\sim1.6 \sigma$) using Cash statistics. Total counts are for the particular source aperture plus the background annular region. The extraction regions and background regions are shown in the stacked image in Figure \ref{fig:stack_image}.}   

\end{deluxetable}

The best-fit parameters for all spectral fits are listed in Table \ref{tab:spec}. We note that the photon index, $\Gamma$, of X1 does not vary much over the course of the observations. However, there does seem to be a marginal increase in absorbing column density during ObsID 16695. X2 shows no deviation from the galactic $N_{\rm H}$ value, but does exhibit a noticeable hardening of $\Gamma$ in ObsID 16696, which then re-softens in ObsID 16697. Without additions to the absorbing column, this implies that during observation 16696, X2 produced intrinsically more hard flux relative to soft flux.  

\citet{Grimes07} found that Haro 11 is surrounded and permeated by an extended soft X-ray component of thermal plasma that extends $\sim6$ kpc outwards from the galaxy. In our analysis, we have specifically chosen apertures that capture the compact regions of X1 and X2 individually, while minimizing the contribution of this extended soft component. For completeness, we attempt a spectral fit using our same absorbed power law model above, with the additive \texttt{xsapec} term (following \citet{Jia11}) for the soft collisionally-ionized thermal component. Because of low counts, fitting this more complex model is only possible for the stacked spectrum of X1. While we obtain a best-fit value for $\Gamma$ which is broadly consistent with our value in Table \ref{tab:speclums} and that of \citet{Grimes07}, the best-fit $N_{\rm H}$ recedes to the galactic value. Moreover, we obtain $kT \sim 20$ keV, which is inconsistent with the value found by \citet{Grimes07} of $kT \sim 0.68$ keV, as well as being substantially above the fitted range. We therefore attempt the same fit, but freeze the value of $kT$ at 0.68 keV. This yields values for both $\Gamma$ and $N_{\rm H}$ which are consistent with our original model values. However, the resulting C-statistic does not represent a statistically significant improvement over the simpler model. We note that the low normalization of the APEC component yields a fitted spectrum virtually indistinguishable from the original fit, indicating that the our aperture is indeed dominated by the hard power law component from the compact source with minimal contribution from the soft thermal component. We attribute this discrepancy to a vastly different choice of apertures from those in \citet{Grimes07} and limitations of fitting this complex model to low count spectra, and so do not discuss it further.

We model the fluxes resulting from the spectral fits using the \texttt{sample\_flux} task. For each spectral fit, we simulate 1000 random sets of parameters based on the multivariate normal distribution using scales defined by the covariance matrix determined from the real data. For our low-count data, $\sim$1000 simulations is necessary to obtain a smooth cumulative distribution function for the randomly drawn parameter values, whose median values are then used in calculating the flux. We quote the resulting median fluxes for X1 and X2 in Table \ref{tab:speclums}. We give the sampled absorbed and unabsorbed fluxes for each observation in the broad, soft, and hard bands. It is important to keep in mind that these values represent upper limits on any XRBs in X1 or X2 since the modelled fluxes are derived from spectra for the entirety of the encircled extraction regions which likely contain ensembles of X-ray objects.  

\begin{table*}
\centering
\caption{Luminosities Derived from Spectral Fitting}
\label{tab:speclums}
\tablewidth{\textwidth} 
\begin{tabular}{lcccccc}
\hline
\hline

\centering
\tabletypesize{\large}
\tablewidth{195.87999pt}
\tablecaption{X1 X-ray Luminosities}

 & \multicolumn{3}{c}{X1} &  \multicolumn{3}{c}{X2}  \\
\hline
ObsID & $L_{\rm broad}$ & $L_{\rm soft}$ & $L_{\rm hard}$  & $L_{\rm broad}$ & $L_{\rm soft}$ & $L_{\rm hard}$\\

 & 0.3$-$8 keV & 0.3$-$2 keV & 2$-$8 keV  & 0.3$-$8 keV & 0.3$-$2 keV & 2$-$8 keV \\

\hline

\vspace{0.2cm}
8175 & 9.8$^{+1.0} _{-0.8}$  & 2.9$^{+0.2} _{-0.2}$  &6.9$^{+1.0} _{-0.8}$  &    3.6$^{+0.4} _{-0.5}$  & 2.0$^{+0.3} _{-0.2}$  &   1.6$^{+0.3} _{-0.3}$ \\ \vspace{0.2cm}
  &   11.1$^{+0.9} _{-0.9}$  & 4.2$^{+0.8} _{-0.7}$  &  7.0$^{+0.9} _{-0.8}$  &    4.4$^{+0.5} _{-0.5}$  & 2.7$^{+0.7} _{-0.4}$  &  1.6$^{+0.4} _{-0.3}$ \\ \vspace{0.15cm}
16695  &   9.7$^{+1.5} _{-1.3}$  & 2.8$^{+0.5} _{-0.4}$  &   6.9$^{+1.4} _{-1.3}$ &   5.4$^{+0.8} _{-0.9}$  & 2.3$^{+0.6} _{-0.4}$  & 3.0$^{+0.8} _{-0.7}$ \\  \vspace{0.2cm}
  &   13.1$^{+2.1} _{-2.3}$  & 6.0$^{+2.6} _{-2.1}$  & 7.0$^{+1.5} _{-1.3}$ &    6.1$^{+0.9} _{-0.8}$  & 3.1$^{+0.6} _{-0.6}$   &  3.0$^{+0.9} _{-0.7}$  \\ \vspace{0.15cm}
16696  &    9.2$^{+1.5} _{-1.4}$  & 2.8$^{+0.5} _{-0.4}$   &   6.4$^{+1.3} _{-1.4}$  &  4.0$^{+0.4} _{-0.4}$  & 1.4$^{+0.3} _{-0.4}$  &  2.3$^{+0.0} _{-0.1}$ \\  \vspace{0.2cm}
  &    10.6$^{+1.2} _{-1.2}$  & 4.1$^{+1.2} _{-0.9}$   &  6.4$^{+1.3} _{-1.2}$ &    4.8$^{+0.2} _{-0.5}$  & 1.8$^{+0.2} _{-0.7}$    &    2.8$^{+0.0} _{-0.0}$ \\  \vspace{0.15cm}
16697  &    8.5$^{+1.2} _{-1.2}$  & 3.1$^{+0.6} _{-0.6}$  &   5.5$^{+1.3} _{-1.0}$  &    6.5$^{+1.0} _{-1.1}$  & 3.5$^{+1.0} _{-0.8}$  & 2.9$^{+0.8} _{-0.6}$  \\  \vspace{0.2cm}
  &   8.9$^{+1.3} _{-1.2}$  & 3.4$^{+0.8} _{-0.7}$  &   5.4$^{+1.4} _{-1.0}$ &    7.8$^{+1.1} _{-1.1}$  & 4.8$^{+1.2} _{-1.0}$ &  3.0$^{+0.8} _{-0.7}$ \\ \vspace{0.15cm}
Stacked  &   9.5$^{+0.6} _{-0.6}$  & 2.9$^{+0.2} _{-0.2}$  & 9.5$^{+0.6} _{-0.6}$   &   4.5$^{+0.3} _{-0.4}$  & 2.2$^{+0.2} _{-0.2}$ & 2.4$^{+0.3} _{-0.3}$\\ \vspace{0.2cm}
  &   10.8$^{+0.6} _{-0.6}$  & 4.1$^{+-6.1} _{-0.5}$ & 6.7$^{+0.6} _{-0.5}$  &  5.1$^{+0.4} _{-0.4}$  & 2.7$^{+0.3} _{-0.3}$ & 2.4$^{+0.3} _{-0.3}$\\

\hline
\end{tabular}

\tablecomments {Spectral fitting-derived luminosities for the regions of X1 and X2. The regions are enclosed in the apertures shown in Figure \ref{fig:stack_image}. Columns 2$-$4 give luminosities for the region of X1, and Columns 5$-$7 give luminosities for the region of X2. For each ObsId, two measurements are given: the absorbed luminosity (top), and the unabsorbed luminosity (bottom), for each energy band. All luminosities are in units of 10$^{40}$ erg s$^{-1}$. The uncertainties are given as the 90\% confidence intervals found using Cash statistics.}  

\end{table*}

\begin{figure*}
	\includegraphics[width=1\textwidth]{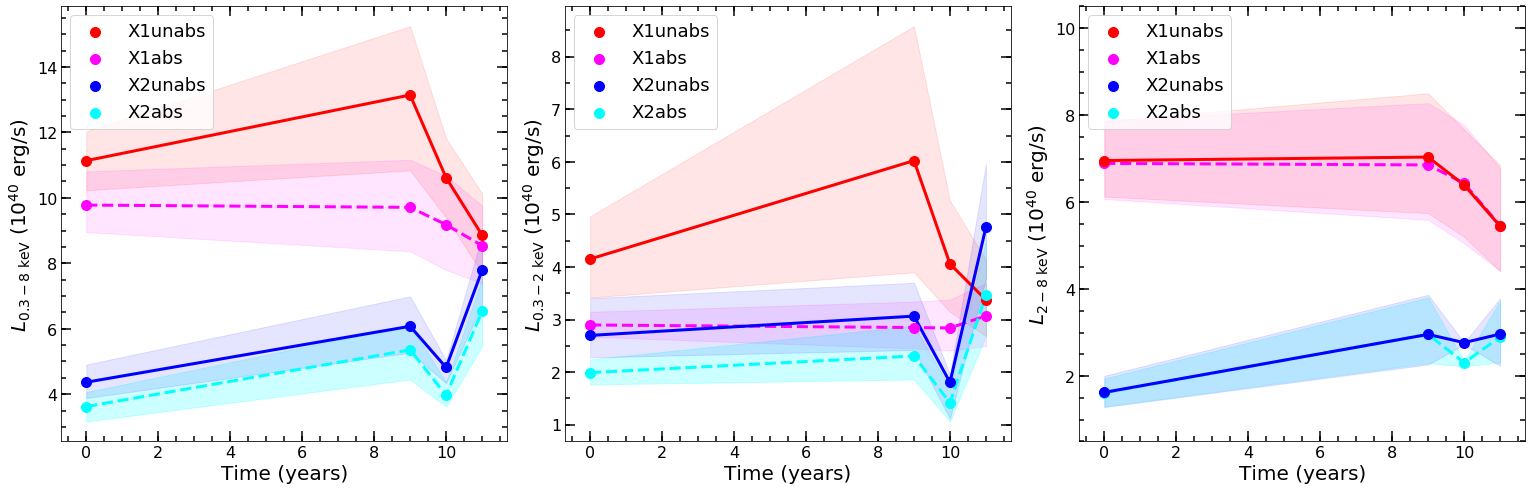}
	\centering
    \caption {X-ray luminosities across the observations, derived from spectral fitting. The x-axis is in years since the first observation (8175), and highlights the short timescale on which the sources vary. {\bf Left:} broad (0.3$-$8 keV) luminosities for the regions of X1 and X2. The absorbed and unabsorbed luminosities are labeled in the legend and displayed as dashed or solid lines, respectively. Error bars are given by the shaded regions and correspond to the 90\% confidence intervals. {\bf Center:} soft (0.3$-$2 keV) luminosities. {\bf Right:} hard (2$-$8 keV) luminosities. The large absorption correction to X1 is unsurprising since it appears to be ensconced in a more extended soft component than X2 in Figure \ref{fig:color_smooth}. X1 appears to be dimming across all energy ranges during the later observations. Meanwhile, X2 experiences an apparent hardening during observation 16696 (at $\sim 10$ years), where its soft flux drops by over 2 orders of magnitude while its hard flux remains fairly constant, followed by a marked flare-up in soft flux. Both regions show some degree of variability, which we discuss further in \S \ref{sec:discussion}. }
    \label{fig:luminosities}
\end{figure*}

In Figure \ref{fig:luminosities}, we plot light curves using fluxes calculated from the spectral fitting results. Immediately obvious is the difference in overall shape between X1 and X2. X1 appears to increase in luminosity before gradually fading at later times across all bands, while X2 exhibits a greater degree of short-term variability. In the soft band, X2 dims during ObsID 16696 (echoed by a harder $\Gamma$), and then becomes brighter than X1 in ObsID 16697. In the hard band, X2 brightens slightly, while X1 gradually dims.

\section{2D Spatial Modelling}
    \label{sec:2D}
 There are noticeable changes in the extended hard band structure of X1 shown in Figure \ref{fig:color_smooth}. This may indicate the presence of multiple X-ray sources. This is contrasted by the more apparent variability of X2, which may contain only a single ULX \citep{Prestwich15}. We explore the possibility of multiple spatial components with fixed positions, but varying fluxes over the course of the observations in both X1 and X2. 

As discussed in \citet{Prestwich15}, the radial profile of hard X-rays in X1 during ObsID 8175 is extended slightly beyond the radial profile of the Point Spread Function (PSF) at that location on the detector. This implies that X1 contains more than just one unresolved point source. However, since radial profiles are one-dimensional cross-sections, information is lost as to the location of a secondary source on the 2D image. We therefore investigate the spatial structure of Haro 11 using 2D modelling in \texttt{Sherpa}. 

We begin by applying the energy-dependent subpixel event repositioning algorithm (EDSER) of \citet{Li04} to all of our evt2 files. This improves the uncertainties of pixel impact positions allowing for accurate subpixel-scale analysis, crucially important for disentangling multiple unresolved point sources on the small angular scales of X1 and X2. We then bin the images to 1/2 the native pixel size ($\sim$0.246\arcsec). We focus on the narrow hard band of 3$-$5 keV so as to avoid the contribution to the flux from diffuse thermal hot gas (kT $\sim$ 1 keV) seen in Figure \ref{fig:color_smooth}. Above $\sim$5 keV the flux drops off rapidly in all observations of X1 and X2, as shown in their spectra. 
For a fair comparison against the results of our spectral modelling, we only model the regions inscribed by the same extraction apertures used above. 

Accurate 2D modelling requires folding in the effects of the detector responses. We simulate PSFs at the individual locations of X1 and X2 using {\it Chandra's} ray tracing software ChaRT \citep{Carter03}, which is then projected onto the detector using MARX \citep{Davis12}. We estimate the 3$-$5 keV flux enclosed within each aperture using \texttt{srcflux} as an input of monochromatic flux at 4 keV. Since the regions contain relatively low fluxes ($\sim$10$^{-6}$ photons cm$^{-2}$ s$^{-1}$), we simulate 50 PSFs at each of the 8 locations. These are then combined within MARX to achieve a more "filled in" stacked PSF image for each region. The PSF images are then binned to the same spatial and energy scales as the evt images, having folded in EDSER within MARX. While it might seem advantageous to our analysis to bin to even smaller pixel sizes, it is unknown how the PSF simulation programs will behave at these smaller scales ({\it e.g.}, 1/8 pixel size) (private correspondence with MARX engineer Hans Moritz G\"unther via CXC Helpdesk). The choice of 50 PSF simulations in each stack is limited by ChaRT; however, even at the 1/2 pixel scaling, the PSF simulations show little variation as expected for observations that are all fairly on-axis.
    
For each region, our model consists of a 2D delta function (point source) and a constant background, convolved with the combined PSF image. Since we are modelling XRBs which are presumably at static locations on the sky, we fix the location of the delta functions during all fits. We determine the best location for a primary source in X1 and in X2 by running \texttt{wavdetect} on the subpixel images. In both cases, we obtain one significant detection whose location is within $\sim$1 (subscaled) pix across the 4 observations. We opt to use the average location for all fits. These positions correspond to on-sky RA, DEC (J200) of 0:36:52.4393, $-$33:33:16.768 for X1 and 0:36:52.6892, $-$33:33:17.048 for X2. Due to Sherpa's limitations when using the \texttt{delta2d} model, positions must be in integer pixel values. The positions above represent the closest whole (subscaled) pixel locations to the average values.

We run the model fitting independently for X1 and X2 (excluding everything outside the aperture regions) using Cash statistics and Neldermead optimization. While both sources could be modelled simultaneously, the extended background components differ locally between X1 and X2. Modelling each source independently allows for better constraints on not only these background components, but also the point source amplitudes. 
    

\begin{figure*}
	\includegraphics[width=\textwidth]{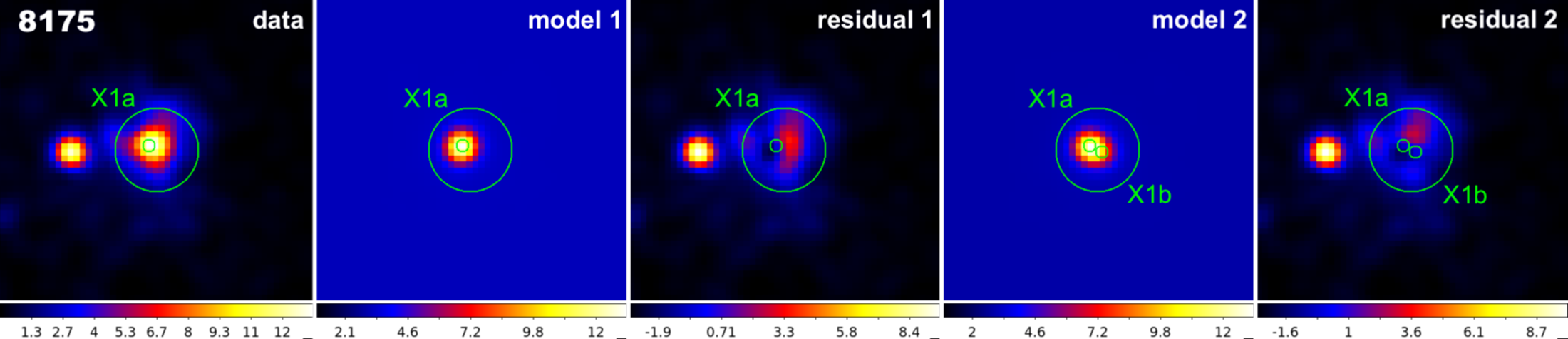}
	\centering
    \caption {2D model fitting example. We show the procedure for obtaining a spatial fit for X1 using the subpixel resolution image of ObsID 8175. Panels from left to right: the real image; the best fit 2D model using a point single source component and constant background convolved with the simulated PSF; the residuals from the first fit; the best fit model using two point source components and constant background convolved with the simulated PSF; and the residuals from the second fit. In all panels, we mark the {\it region} of X1 (green circle) to be exclusively fit. The smaller green circles denote the fixed positions of the \texttt{delta2d} functions used to fit the sources determined via \texttt{wav\_detect}. We start by fitting X1a alone, and then X1a along with X1b. The images are at half pixel scaling (0.249\arcsec) and are in the narrow hard band of 3$-$5 keV. The images are individually scaled given by the colourbar at their bases in units of total counts over the exposure. As before, up is North, and left is East. This procedure is done for X1 and X2 individually, for each ObsID.}
    \label{fig:2D_procedural}
\end{figure*}

\begin{figure}
	\includegraphics[width=1\columnwidth]{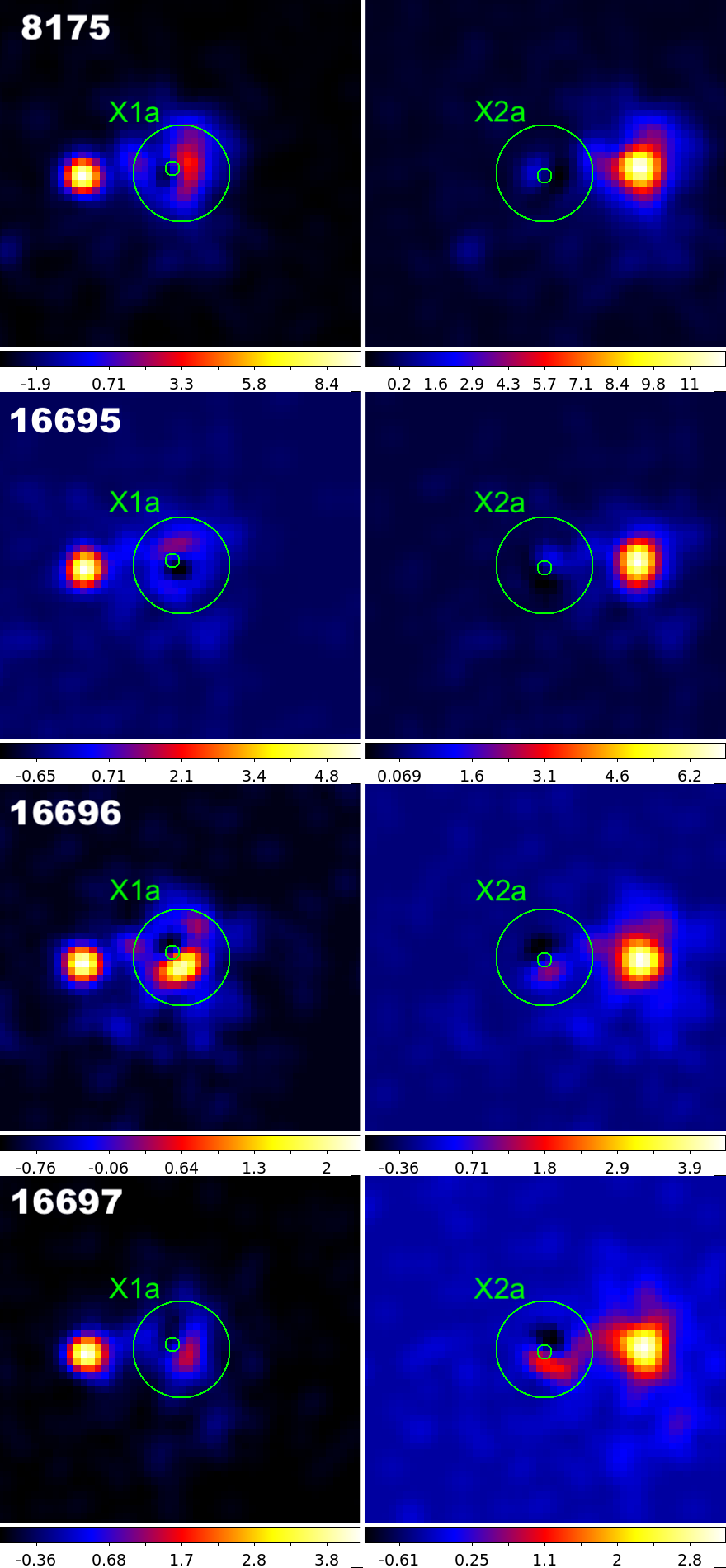}
	\centering
    \caption {2D fitting residuals for the single source models. For each row (ObsID), residuals from the 2D fitting procedure are shown for X1 in the left column, and X2 in the right column. The residuals are a result of the best fit model subtracted from the real image. In all panels, we mark the {\it region} of X1 (green circle) to be exclusively fit. The smaller green circles denote the fixed positions of the \texttt{delta2d} functions used to fit the sources. The reader can see what the unfitted data image for X1 looks like in the corresponding panel for fits to X2, and vice versa, since nothing outside the aperture is considered during the fitting. The images are at half pixel scaling (0.249\arcsec) and are in the narrow hard band of 3$-$5 keV. The images are individually scaled given by the colourbar at their bases in units of total counts.}
    \label{fig:2D1}
\end{figure}

In Figure \ref{fig:2D_procedural}, we show the workflow of the 2D fitting procedure, focusing on the region of X1 during observation 8175 as an example. The first three panels from the left show the routine for fitting a model consisting of a single point source (whose position is shown with the smaller green circle) plus constant background, convolved with the PSF. All images are of the narrow hard band, at half pixel scaling. It is immediately obvious from the residuals that a single point source smeared by the PSF cannot account for the rich extended structure of X1. Perhaps more interesting is that the apparent location of the highest concentrations of residuals appears to change between observations. We show the resulting residual images for each fit in Figure \ref{fig:2D1}, where X1 and X2 are separately fit using the single point source model. By eye, there seems to be a prominent blob of residuals to the West of the primary source of X1 in 8175, and a Southwest blob in both 16696 and 16697. This hints that there is at least one secondary component in the X1 region. Modelling of X2 using a single point source leaves fewer residuals. However, a faint Southern component appears in 16696, and persists through 16697. While the coincidence of the southwestern residuals in both X1 and X2 might at first indicate an issue with the astrometry, we note that the residuals in observation 16697 differ in orientation between X1 and X2, where the residuals for X1 are more westerly, and those for X2 are more southerly, suggesting that changes in the apparent structure are not coordinated. We also note that the difference between the detected primary source positions of X1 and X2 for ObsID 16696 do not differ from the positions found on the stacked image by a systematic amount. While still less than 1 sub-sampled pixel, the positional difference for X1 amounts to $\sim 0.2\arcsec$, and $\sim 0.07\arcsec$ for X2. As can be seen in the residual images, the blobs account for a much wider area than could be accounted for by these small positional offsets. We therefore suggest that the coincidence of Southwestern residuals is due to intrinsic variations in X1 and X2.  

Table \ref{tab:2Dlums} lists the best-fit model components for the single point source models. Columns two and three give the values for the single source component for X1a, taken to be the primary source in the region of X1, and likewise wise for X2a. We also give the total value for the model, {\it i.e.}, the point source plus the extended background component. To compute rough estimates of the fluxes encapsulated by these components, we normalize the counts by the exposure map value at the fixed pixel locations and convert to ergs assuming an average photon energy of 4 keV. We note that while the constant background component is spatially extended across the $\sim$9 arcsec$^{2}$ area of the fitted region, the exposure map does not vary appreciably on this scale for our observations close to the detector aim point. 


\begin{table*}
\centering
\caption{2D Modelling Results}
\label{tab:2Dlums}
\tablewidth{\textwidth} 
\begin{tabular}{l||cccc||cccccc}
\hline
\hline

\centering
\tabletypesize{\large}
\tablewidth{195.87999pt}
\tablecaption{X1 X-ray Luminosities}

 & \multicolumn{2}{c}{X1a}  &   \multicolumn{2}{c||}{Total model 1} & \multicolumn{2}{c}{X1a} &  \multicolumn{2}{c}{X1b} &   \multicolumn{2}{c}{Total model 2}\\

ObsID & cts & $L_{\rm X}$ & cts & $L_{\rm X}$  & cts & $L_{\rm X}$ & cts & $L_{\rm X}$ &cts & $L_{\rm X}$\\

\hline

\vspace{0.2cm}
8175  &   232.2$^{+35.2} _{-33.1}$  & 5.61$^{+0.85} _{-0.8}$  & 700.74$^{+54.6} _{-51.8}$  &16.93$^{+1.32} _{-1.25}$  &  188.7$^{+34.4} _{-32.1}$  & 4.56$^{+0.83} _{-0.78}$  & 90.7$^{+29.5} _{-27.0}$  & 2.19$^{+0.71} _{-0.65}$  & 703.35$^{+61.4} _{-57.5}$  &16.99$^{+1.48} _{-1.39}$ \\\vspace{0.2cm}

16695  &   165.7$^{+27.0} _{-25.0}$  & 8.95$^{+1.46} _{-1.35}$  & 274.96$^{+35.5} _{-32.6}$  &14.86$^{+1.92} _{-1.76}$  &  161.4$^{+27.5} _{-25.5}$  & 8.72$^{+1.49} _{-1.38}$  & 8.6$^{+15.2} _{-12.3}$  & 0.46$^{+0.82} _{-0.66}$  & 275.23$^{+39.3} _{-35.5}$  &14.87$^{+2.12} _{-1.92}$  \\\vspace{0.2cm}
16696  &   71.8$^{+20.8} _{-18.7}$  & 3.89$^{+1.13} _{-1.01}$  & 226.9$^{+32.7} _{-29.8}$  &12.29$^{+1.77} _{-1.61}$ &  43.8$^{+19.3} _{-16.7}$  & 2.37$^{+1.05} _{-0.91}$  & 49.6$^{+19.0} _{-16.7}$  & 2.69$^{+1.03} _{-0.91}$  & 227.43$^{+36.2} _{-32.3}$  &12.33$^{+1.96} _{-1.75}$  \\\vspace{0.2cm}
16697  &  53.9$^{+18.3} _{-16.1}$  & 3.05$^{+1.04} _{-0.91}$  & 186.25$^{+29.4} _{-26.5}$  &10.56$^{+1.66} _{-1.5}$  &   34.8$^{+17.0} _{-14.6}$  & 1.97$^{+0.96} _{-0.82}$  & 43.7$^{+18.0} _{-15.6}$  & 2.48$^{+1.02} _{-0.88}$  & 186.79$^{+33.1} _{-29.2}$  &10.59$^{+1.88} _{-1.65}$   \\
\hline
\hline
 & \multicolumn{2}{c}{X2a}  &   \multicolumn{2}{c||}{Total model 1} & \multicolumn{2}{c}{X2a} &  \multicolumn{2}{c}{X2b} &   \multicolumn{2}{c}{Total model 2}\\
\hline
\vspace{0.2cm}
8175  &   263.7$^{+30.9} _{-29.1}$  & 6.39$^{+0.75} _{-0.71}$  & 378.1$^{+38.6} _{-35.8}$  &9.16$^{+0.94} _{-0.87}$ &   260.7$^{+31.8} _{-29.9}$  & 6.32$^{+0.77} _{-0.72}$  & 4.7$^{+14.4} _{-11.6}$  & 0.11$^{+0.35} _{-0.28}$  & 378.19$^{+42.0} _{-38.3}$  &9.17$^{+1.02} _{-0.93}$ \\\vspace{0.2cm}
16695  &   127.3$^{+21.2} _{-19.4}$  & 6.94$^{+1.15} _{-1.06}$  & 163.19$^{+25.1} _{-22.4}$  &8.89$^{+1.37} _{-1.22}$  &  127.3$^{+21.2} _{-19.4}$  & 6.94$^{+1.15} _{-1.06}$  & 0.0$^{+5.7} _{-0.0}$  & 0.0$^{+0.31} _{-0.0}$  & 163.19$^{+25.8} _{-22.4}$  &8.89$^{+1.4} _{-1.22}$\\\vspace{0.2cm}
16696  &  66.3$^{+16.5} _{-14.6}$  & 3.6$^{+0.89} _{-0.79}$  & 106.33$^{+21.3} _{-18.6}$  &5.77$^{+1.16} _{-1.01}$  &  42.4$^{+15.9} _{-13.6}$  & 2.3$^{+0.86} _{-0.74}$  & 33.6$^{+15.4} _{-13.2}$  & 1.82$^{+0.84} _{-0.72}$  & 106.96$^{+25.3} _{-21.5}$  &5.8$^{+1.37} _{-1.17}$  \\\vspace{0.2cm}
16697  &  107.4$^{+20.8} _{-18.9}$  & 6.11$^{+1.18} _{-1.08}$  & 161.76$^{+26.5} _{-23.7}$  &9.2$^{+1.51} _{-1.35}$  &  87.2$^{+20.4} _{-18.4}$  & 4.96$^{+1.16} _{-1.04}$  & 41.3$^{+17.2} _{-14.7}$  & 2.35$^{+0.98} _{-0.84}$  & 163.31$^{+30.5} _{-26.7}$  &9.28$^{+1.74} _{-1.52}$  \\

\hline
\end{tabular}

\tablecomments {2D modelling-derived luminosities. The fitting results for the regions of X1 and X2 are separated by the horizontal middle line. The top half gives the fit results of X1, and the bottom half gives the results for X2. Columns 2$-$5 give the best-fit values for the model comprised of a single source component (X1a or X2a) plus the constant background. Columns 6$-$11 give the best-fit values for the model comprised of 2 source components (where a is the primary and b is the secondary) plus the constant background. In all cases, {\it Total} gives the full value of the model fitted within the aperture ({\it i.e.}, Total model 1 gives the value for the single source and background), which is shown to be consistent between the one and  two-source models. Values for each model component's amplitude are given in counts. Hard band luminosities $L_{\rm X}$ are in units of 10$^{40}$ erg s$^{-1}$. All 2D model fitting is done on images in the hard band (3$-$5 keV) at a scale of 1/2 the native pixel size (0.246\arcsec).}  
\end{table*}

To test whether the level of residuals indicates a statistically significant secondary source, we run \texttt{wavdetect} on the residual images. Once again for both X1 and X2, we detect 1 source per region. However, in both cases the 4 detections do not agree unanimously. For X1, all observations except 16695 yield detections consistent to within $\sim$1 (subscaled) pixel. For X2, ObsIDs 16696 and 16697 yield similarly consistent detections, while 8175 and 16695 yield a separate, though slightly less consistent (within $\sim$1.75 (subscaled) pixels) detection. We adopt the more consistent detections as the most promising secondary source locations, adding them to the previous 2D model with fixed positions. These positions correspond to on-sky RA, DEC (J200) of 0:36:52.4056, $-$33:33:17.033 for X1b and 0:36:52.6811, $-$33:33:17.525 for X2b. The only additional requirement imposed is that the source amplitudes cannot go negative (a possible modelling outcome for two closely separated sources), which would imply non-physical negative fluxes. 

\begin{figure}
	\includegraphics[width=1\columnwidth]{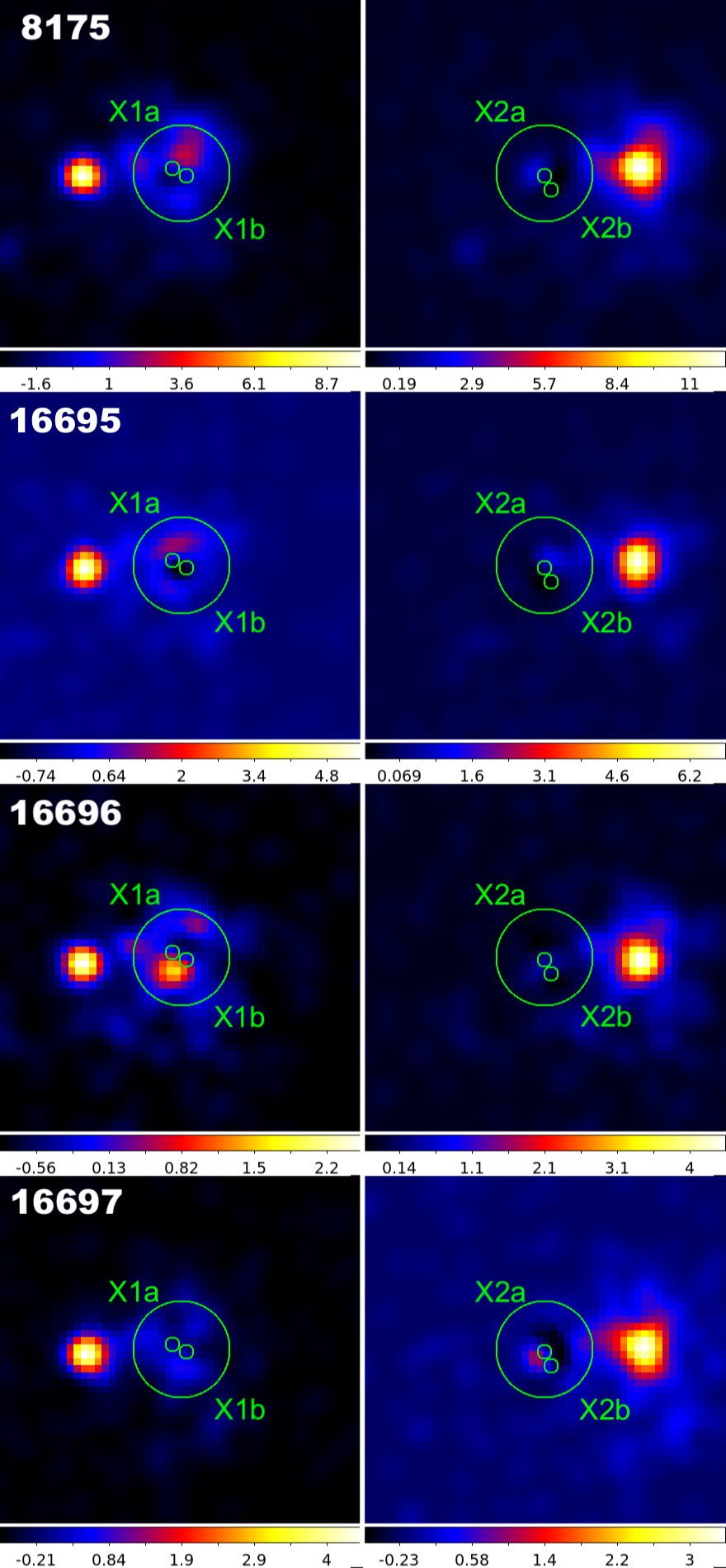}
	\centering
    \caption {Same as \ref{fig:2D1}, but for models containing 2 source components. In both X1 and X2, the primary component (denoted as `a') is towards the Northeast, and the secondary component (denoted as `b') is towards the Southwest.}
    \label{fig:2D2}
\end{figure}

We run the fitting routine again on the original data image, resulting in the residual images in Figure \ref{fig:2D2}. In both cases of X1 and X2, the secondary source lies towards the Southwest, although to differing degrees. The two-source model components and their estimated luminosities are given in Table \ref{tab:2Dlums} columns 6$-$11. The more complex model yields noticeably better residuals for X1 across the board. In X2, the secondary component (X2b) contributes nearly no counts in the first two observations; however, the latter two observations see a marked increase of both secondary components. 

While the more complex 2D models appear to yield better fits to the data by eye, we check that these improvements are statistically significant. The Cash statistic is a log-likelihood function, so we compare the resulting fit statistics to conduct a simple likelihood ratio (LR) test between the more complex 2-source component model and the nested 1-source component (null) model. We note that there is only one additional degree of freedom between the two models (amplitude of component b). In all cases for X1, LR $>$ 4, indicating an improvement in the fit at the 95\% confidence level. X2 yields similar LRs, except in the case of 16695 where LR $\sim$ 3, indicating an improvement at only the $\gtrsim$90\% confidence level. This is not surprising given the disagreement in secondary source positions for X2. However in the context of variability, we note that for the lower flux secondary sources these particular observations may indicate periods where the XRB is at a decreased brightness such that it is not adding a perceptible amount of flux to the ensemble as a whole. 

\begin{figure}
	\includegraphics[width=\columnwidth]{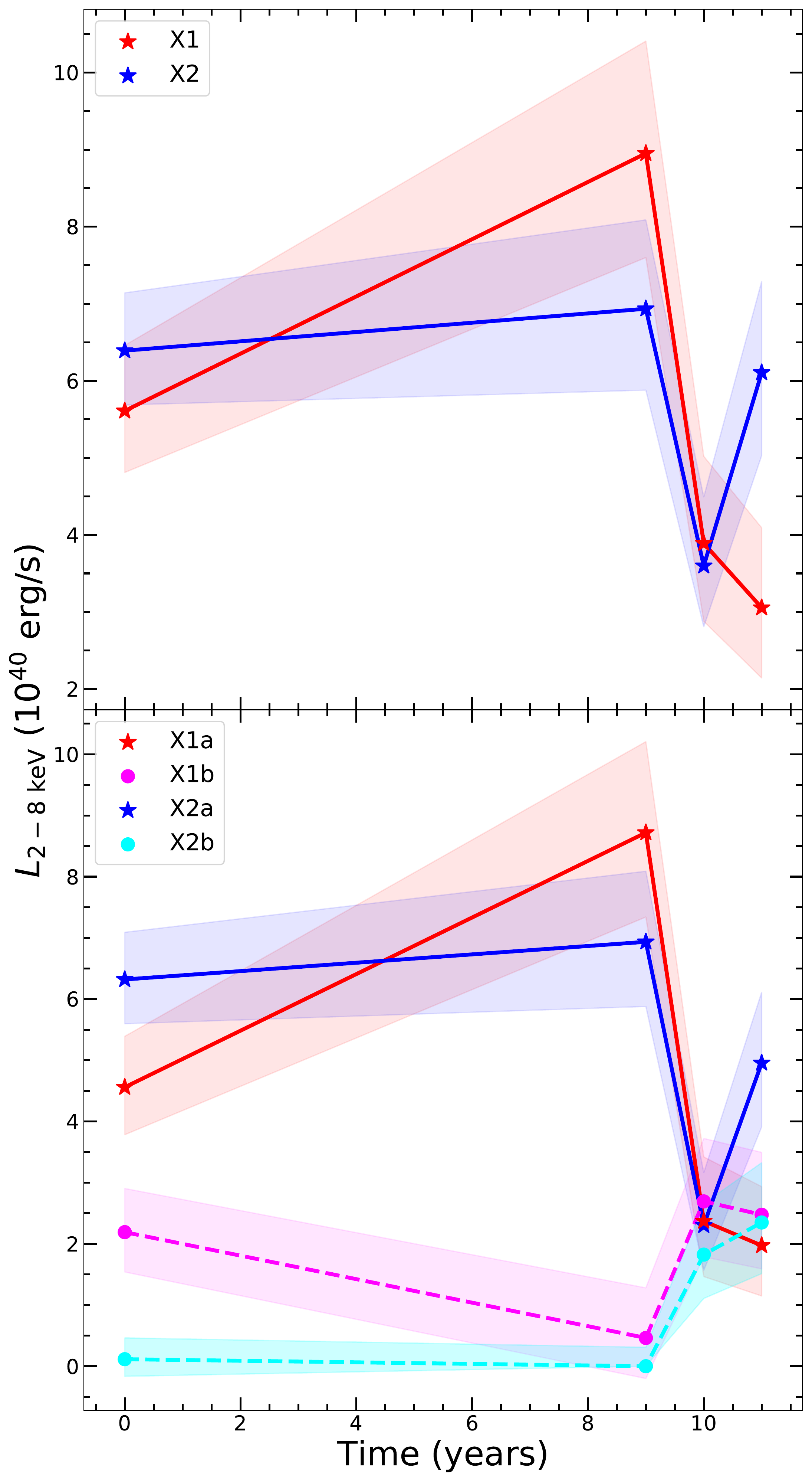}
    \caption {{\bf Top:} Lightcurves of the modelled sources in the 1-component model fits. The x-axis is in years since the first observation (8175), and highlights the short timescale on which the sources vary. The error bars are the 90\% confidence limits. Both sources show some degree of variability. Perhaps most interesting is that X2 is brighter in the hard band than X1 in the final observation, differing from the result obtained via spectral fitting over the entirety of the extracted regions. This is not surprising since X2 is more spatially compact than X1, and the single point source model is only capturing a portion of its extended structure. {\bf Bottom:} same as above, but for models containing 2 point sources. For a given pair, the primary source (a) is shown as a solid line and marked with stars, while the secondary source is plotted with a dashed line and marked with circles. While the overall shape of the primary source components is the same as for the 1-component model, their luminosities are slightly reduced. Here we also see that the combined luminosity of the X1 ensemble starts out brighter than the X2 ensemble, but becomes fainter or comparable at later times.}
    \label{fig:model_Lx}
\end{figure}

The trends for the various model fits are more clearly illustrated by the lightcurves shown in Figure \ref{fig:model_Lx}. Both regions clearly have a dominant source, whose luminosities are modestly affected by the addition of a secondary component. In the lower panel, we see that the primary components dominate the hard X-ray emissions of both regions during the earlier observations. It is interesting that X1b overtakes X1a during the last two observations, becoming the dominant source in the region. We discuss this interplay and possible physical interpretations in \S\S \ref{sec:X1}.

\section{Discussion}
    \label{sec:discussion}
Our extended program of observations has allowed us to uncover a more intricate picture of the complex Haro 11 system. We begin by addressing whether either X-ray region exhibits a higher degree of variability than what might be expected for similar ensembles of X-ray sources vis-\`{a}-vis M82. We then combine the evidence revealed through our analyses to offer revised pictures of both X1 and X2, which strengthen some of our previous conclusions.

\subsection{Variability and Number of Sources}
    \label{sec:variability}
    
In \citet{Prestwich15}, we compared the region of Haro 11 X1 to the young starburst region of M82. This nearby (d$\sim$3.6 Mpc \citep{Freedman94}) galaxy has been observed using $Chandra$ nine times between the years 1999 and 2007. \citet{Chiang11} detected 58 X-ray sources (with $L_{\rm X} > 10^{37}$ erg s$^{-1}$) within the $D_{25}$ optical isophote of M82 ($\sim11.6\ \times\ 5.6$\ arcmin$^{2}$ corresponding to $\sim 12\ \times\ 6$ kpc$^{2}$), although 25 of those sources reside in the central 1 $\times$ 1 arcmin$^{2}$ region. They find that 26 of their sources show variability in flux on the scale of days to years, with 17 sources exhibiting a maximum flux variation of factors $> 3$. However, the central region is dominated a bright ($\sim$ 10$^{41}$ erg s$^{-1}$ \citep{Kaaret09}) X-ray source, M82 X-1, which has been shown to vary in flux by a factor of $\sim$8 \citep{Kaaret01, Matsumoto01}. This source is thought to be a highly luminous ULX in the hard state, likely powered by an IMBH \citep{Kaaret09}. 

Haro 11 is much farther away than M82 and so individual sources cannot be resolved. It is therefore useful to draw comparisons to M82, where the total X-ray output is known to be a composite of many resolved XRBs. Like M82, the distribution of sources in the region of Haro 11 X1 is spatially compact. The inner annular radius shown in Figure \ref{fig:stack_image} corresponds to 6.5 $\times$ 6.5 arcsec$^{2} \sim 5.2 \times 5.2 $\ kpc$^{2}$ at a distance of 84 Mpc, similar to the region analyzed by \citet{Chiang11}. While the aperture used in our spatial analysis corresponds to almost double the angular size of the central starburst core of M82 (see \citet{Prestwich15} Figure 7 for a size comparison), the point sources in our spatial modelling are much more closely separated. The distance between the {\it individual} source components is $<$1 kpc (distance between the positions of X1a and b : d $\sim$ 300 pc).

The extent of the region of X2 is even more spatially compact than X1, although the point source components in our spatial modelling are separated by approximately the same amount as X1a and X1b. Since Haro 11 is a metal-poor system \citep{James13} similar to Lyman Break galaxies that would supposedly have an excess of ULXs \citep{BasuZych13}, it seems plausible that there would be ongoing BH mergers near the dynamical center of the galaxy. In fact, \citet{Brorby14} found that the X-ray luminosity function (XLF) of low-metallicity BCDs suggests a significant enhancement of High Mass XRBs (HMXBs). Similarly, \citet{BasuZych16} found that the LBAs Haro 11 and VV 114 contain $\sim$4 times as many ULXs ($L_{\rm X} > 10^{41}$ erg s$^{-1}$) than would be expected based on their star formation rates (SFRs), owing to a shallower slope at the bright end of their XLFs. Star formation in the central starburst of M82 is thought to have been going on for $\sim$60 Myr \citep{Gallagher99}, so it is unsurprising that it contains multiple XRBs. With these traits in mind, we take M82 to be a reasonable local comparison to Haro 11. 

The concentration of XRBs in Haro 11 is also not surprising when compared to other nearby starburst galaxies with irregular morphologies due to ongoing galactic mergers. In NGC 4449, \citet{Rangelov11} identified 7 HMXBs within a radius of $\sim$1.2 kpc, similar to the radius encompassing both Haro 11 X1 and X2 ($\sim$1.36 kpc). They concluded that the XRBs are likely dominated by black holes, and are coincident or within 200 pc of $\lesssim$8 Myr age star clusters. The most luminous of these XRBs is only $L_{\rm 0.3-8kev}\sim 3 \times10^{38}$ erg s$^{-1}$; however one of the additional XRBs detected outside this region approaches $\sim 9\times10^{38}$ erg s$^{-1}$. More luminous XRBs were detected by \citet{Zezas06} in the Antennae Galaxies, who found 10 ULXs with $L_{\rm 0.3-7kev}\geq 10^{39}$ erg s$^{-1}$, in addition to dozens of XRBs with $L_{\rm X}\geq 10^{37}$ erg s$^{-1}$. Of these, \citet{Rangelov12} confirmed that 22 of the XRBs are coincident with star clusters, 14 of which had an age of $\lesssim$6 Myr. The coincidence of XRBs with super star clusters seems to be a common feature of these solar-metallicity irregular galaxies; however, the lower metallicity of Haro 11 may help to explain the highly luminous sources found in the regions  X1 and X2. 

To assess variability in Haro 11, we calculate the maximum variations in flux for the {\it whole regions} of X1 and X2 between observations given from the values in Table \ref{tab:2Dlums}. We find that over the course of our four observations, neither region exhibits a broadband flux variation in excess of $\sim$77\%, with X2 being the more variable source. We note that the percent changes are calculated for the regions of X1 and X2 designated by the apertures in Figure \ref{fig:stack_image}. While our spatial analysis was able to detect a pair of sources in each region, they likely contain additional XRBs below the threshhold of detection, which would also be contributing to the X-ray emission of the regions X1 and X2. The flux variations for the individual sources in the lower panel of Figure \ref{fig:model_Lx} are similiar to the variability seen from XRBs in M82, with maximum changes in flux by factors of $\sim$3$-$4.5. We therefore conclude that the X-ray regions in Haro 11 are no more variable than the X-ray sources in M82.

Our spatial analysis above suggests that the X-ray emission in Knot B of Haro 11 cannot be fully explained by a singular source, contrary to our assessment in \citet{Prestwich15}. Instead, the emission of the X1 region is likely due to an ensemble of several accreting sources that are in such close proximity that they cannot be individually resolved by $Chandra$. Our best-fit model depicts the region as two sources, where the more dominant source is readily targeted by \texttt{wavdetect}, and a secondary source is consistently found at the same location in the residuals. It may in fact be that there are additional sources contributing to the extended hard X-ray structure of the region to a lesser degree as a result of source blending \citep{BasuZych16}; however, we do not attempt to estimate their positions or fluxes at the risk of over-modelling the scant remaining counts in the residual images.

Our spatial analysis of X2 also suggests that in at least several observations, multiple point sources yield a better fit to the data than an individual source. We adopt the same convention as with the complex in X1, denoting the primary component as X2a, and the secondary component as X2b.

\subsection{Optical Signatures of AGN Activity}
    \label{sec:BPT}

Given the best-fit luminosities found above, we consider the possibility that the region of X1 contains an AGN. The X-ray luminosities estimated above for X1 as an extended region, as well as the individual components X1a and X1b, show evidence of AGN-scale accretion, albeit in the range of low luminosity AGN (LLAGN) ($L_{\rm X} \sim 10^{40}$erg s$^{-1}$; for a review, see \citet{Ho08} and references therein). A dearth of LLAGN have been found in the local universe using $Chandra$ by \citet{She17}. And given that Haro 11 appears to be in the process of a galactic merger \citep{Ostlin15}, any detected LLAGN might be a precursor to a central SMBH after coalescence. 

\begin{figure}
	\includegraphics[width=\columnwidth]{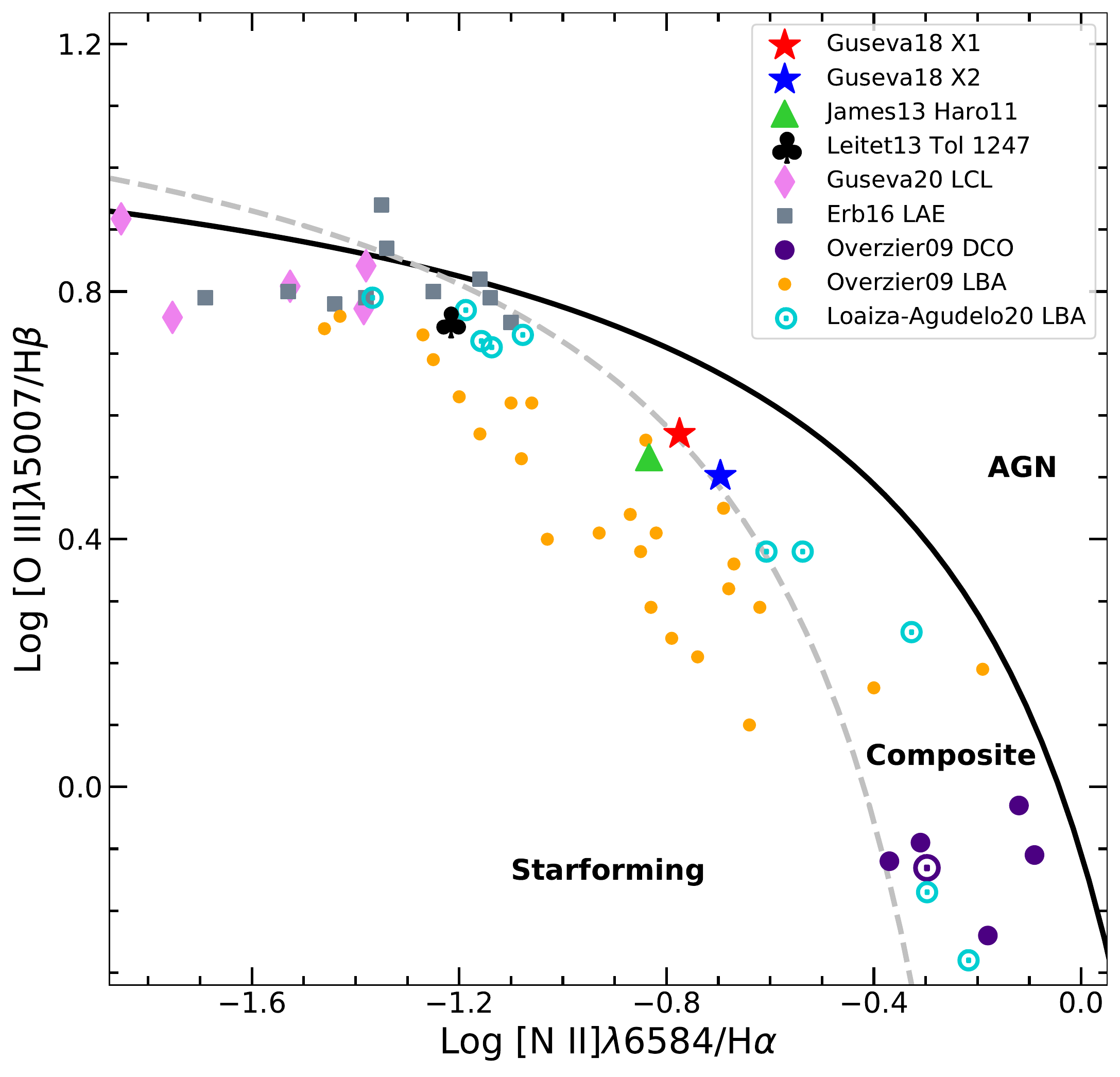}
    \caption{BPT diagram of Haro 11. We compute and plot the line ratios for the regions X1 and X2 based on the extinction-corrected line fluxes from \citet{Guseva12}. Both Haro 11 X1 and X2 (given by stars) fall within the AGN/composite region, which is bounded by the theoretical "maximum starburst line" of \citet{Kewley01} (solid line). However, both fall closer to the cusp of the demarcation between starburst galaxies and AGN of \citet{Kauffmann03} (dashed line). We also show the ratios derived from the de-reddened line intensities of the integrated spectrum from \citet{James13}. It is important to note that their integrated spectrum covers not only the regions of X1 and X2 (Knots B and C), but also Knot A, whose high level of star formation (but lack of an ionizing X-ray source (see Figure \ref{fig:Lya})) likely pushes the integrated ratios into the pure star-forming realm of the diagram. This is consistent with the placement by \citet{Leitet13} which considers the galaxy on the whole and not the individual Knots. We also show the placement of Tol 1247 from \citet{Leitet13} for comparison. While \citet{Kaaret17} do not rule out an AGN contribution in Tol 1247, the optical line ratios for Haro 11 X1 and X2 appear more consistent with a mix of AGN and star-forming regions. We note that the error bars for these two points are smaller than the markers, and thus do not cross the lower demarcation. The remaining samples drawn from the literature are briefly named in the legend, and represent a wide range of Lyman emitters and Lyman Break Analogs, some of which are at moderate redshifts ($z \gtrsim 2$), and which have extreme properties that place them near or within the composite region. In particular, we denote the most robust LLAGN detection by \citet{Alexandroff12} (J0921+4509), from the original \citet{Overzier09} sample of LBAs, with the largest purple "sun" symbol in the bottom right corner of the plot as a useful comparison object to Haro 11.}
    \label{fig:BPT}
\end{figure}

We draw on literature values of optical emission line fluxes to construct the BPT line ratios \citep{Baldwin81}. The relative ratios of the nebular emission lines ([\ion{O}{iii}]/H$\beta$ and \ion{N}{ii}/H$\alpha$) are diagnostic of the underlying ionization mechanism. Recent VLT/X-shooter observations by \citet{Guseva12} of Haro 11 allow for a dissection of the three individual star-forming Knots. We compute the line ratios of Knots B (X1) and C (X2) from their quoted extinction-corrected line fluxes, and plot them in Figure \ref{fig:BPT}. We also plot the classic demarcations that separate AGN and pure star-forming regions. Ratios above the theoretically modelled cutoff for the maximum contribution to ionization by starburst regions strongly suggest AGN origins \citep{Kewley01}. A lower cutoff gives the observed bounds between sequences of AGN and starburst galaxies in these ionization ratios \citep{Kauffmann03}. We find that the both X1 and X2 fall in between these two extremes, suggesting a mix of AGN and starburst contributions to the ionization field. We therefore revise our previous conclusion from \citet{Prestwich15}, and suggest that both X1 and X2 do have some optical indications of AGN activity.

This result is contrary to previous optical studies of Haro 11, which have relied on line ratios based on a single spectrum for the entirety of Haro 11 \citep{Vader93, Bergvall00, Leitet13}. This is also seen in the case of integral field unit (IFU) observations of Haro 11 by \citet{James13}. We use their de-reddened line fluxes for the integrated spectrum to plot the ratios for Haro 11 {\it as a whole} in Figure \ref{fig:BPT}. While the difference between the whole galaxy and the individual components is small, it does push the ratios securely into the realm of pure star-formation. This is not surprising given that the integrated spectrum includes Knot A, which has a high level of star formation but no obvious X-ray emission (and thus, likely no AGN), as shown in Figures \ref{fig:Lya} and \ref{fig:color_smooth}.  Indeed, a similar effect is seen in the sample of optically-selected AGN in the dwarf galaxy sample of \citet{Mezcua20}. They find that while 37 of their galaxies exhibit AGN optical signatures in sections of their spatially-resolved IFU spectra, single-fibre spectra fail to detect the signatures in 23 of the same galaxies. Interestingly, these 23 additional AGN candidates have bolometric luminosities lower than even just the broad band X-ray luminosities of our sample ($\lesssim$ 10$^{40}$erg s$^{-1}$) \citep{Mezcua20}. 

As noted by \citet{She17}, optical detections of LLAGN are routinely plagued by additional signatures from star formation and starlight from the host galaxy in general, conflating the emission from the nuclear region.
Even in the X-ray regime, starburst regions can create shocks that boost electrons to relativistic energies leading to synchrotron radiation and inverse Compton upscattering of IR photons off these electrons. The residual hard X-ray background of our 2D fitting, as well as the diffuse soft flux permeating the X1 region, are likely dominated by these star-formation effects instigated by the relatively young starburst in Knot B ($\sim$3.5$-$4.3 Myr \citep{Adamo10, James13}). We note that the extended background model component (Total 1 - X1a) captures more counts in observation 8175 than the others (roughly triple), because of the longer exposure time. So longer exposures do not necessarily draw out the putative compact source(s), but instead expose the complicated extended emissions from the starburst that may mask such sources.  We do not see the same scale of counts in the background component of X2, possibly because the star formation in X2 has been going on for a longer time ($\sim$7.4$-$10 Myr, \citet{James13, Adamo10}). 

Based on a sample of 8 LBAs, \citet{Overzier08} suggest that LBAs are usually hosts of super starbursts triggered by gas-rich, low-mass mergers. Starbursts with such extreme properties might even create ionization conditions that mimic optical line ratios seen in the composite region of the BPT diagram \citep{Overzier09}. As such, masking of LLAGN signatures by starbursts may be a common feature of LBAs. \citep{Alexandroff12} further tested this by searching for compact radio cores in the LBAs of \citet{Overzier09} that have BPT composite classification (shown in Figure \ref{fig:BPT} as the large purple circles). Of the four composite LBAs observed at 1.7 GHz, only one (J0921+4509) was detected to have a compact core, along with radio luminosity above what would be expected for star-formation alone, consistent with a possible LLAGN (shown as the open purple circle) \citep{Alexandroff12}. The same set of LBAs was also observed in the X-ray by \citet{Jia11}, who found that all of the LBAs, especially the source detected in radio, have X-ray luminosities above the empirical X-ray-Far-Infrared luminosity relation. With X-ray luminosities of $10^{41}\lesssim L_{\rm 2-10kev}\lesssim 10^{42}$ erg s$^{-1}$ these LBAs show evidence of containing obscured LLAGN that might serve as analogs to hint at conditions of black hole growth in the early universe \citep{Jia11}, although J0921+4509 seems to currently be the most promising candidate. 

We show several other samples from the literature as a comparison in Figure \ref{fig:BPT}. In addition to the local LBAs from \citet{Overzier09} and the composites containing high SFR Dominant Compact Objects (DCOs), we show metal-poor LBAs from \citet{LoaizaAgudelo20}, and Lyman continuum leakers (LCLs) from \citet{Guseva20}. We also show Lyman $\alpha$ emitters (LAEs) from \citet{Erb16}. The latter are objects at higher redshift ($z \sim 2$), while the rest are fairly nearby ($z \lesssim 0.3$). It is interesting that Haro 11 occupies a rather isolated portion of the BPT diagram, especially compared to other Lyman emission sources. Taken together, they represent a diverse assortment of extreme starbursting galaxies which hedge the AGN/composite region of the BPT diagram.

Complicating the picture, it has been seen in samples of high redshift starforming galaxies (SFGs) that there is an offset {\it towards} the composite region when compared against similar SFGs at $z \sim 0$ \citep{Steidel14, Shapley15, Guseva20, Bian20}. This offset is thought to be due mainly to a higher ionization parameter and harder ionizing radiation field in the higher-$z$ SFGs, which is driven by higher SFR (and thus starburst age-dependent)  \citep{Bian20}, and not necessarily caused by an AGN. However, extremely metal-poor galaxies with high ionization parameters have also been shown to deviate from the trends of both low and high $z$ SFGs, in a direction {\it away} from the composite region, suggesting that location on the BPT diagram may also be metallicity-dependent \citep{Izotov21}. While not shown in Figure \ref{fig:BPT}, we refer the reader to Figure 1 of \citet{Bian20} to see how the offset in their stacked $z \sim 2$ analogue sample crosses into the composite region around emission line ratio values of log [\ion{O}{iii}]/H$\beta$ $\sim$ 0.45 and log \ion{N}{ii}/H$\alpha$ $\sim$ -0.65. This is close to the placement of Haro 11 X2, and X1 to a lesser degree. \citet{Bian20} note that the properties of their analogue galaxies resemble those of high redshift LAEs, especially higher ionization parameter, as seen in the sample from \citet{Erb16}. This is also seen in the low-$z$ sample of LCLs from \citet{Guseva20} and the low-$z$ LBAs from \citet{LoaizaAgudelo20}. As a local LAE with fairly young star clusters, Haro 11 might be interpreted as a high-$z$ analogue whose placement on the BPT diagram is influenced by a SFR-dependent ionization potential, especially when taking the system as whole (including Knot A). However, as found by \citet{Guseva12}, the specific SFR of Haro 11 Knot C is over a factor of 10 lower than in Knot B. As well, the low metallicity of Haro 11 might also nudge it towards the SFG region of the BPT diagram, although this should be minimal considering the metallicity of Haro 11 is only marginally less than that seen in the samples mentioned above at 7.8 $\leq$ 12 + log O/H $\leq$ 8.2 \citep{James13} compared to the extremely metal-poor galaxies in \citet{Izotov21} (12 + log O/H $\sim$ 6.0$-$7.25).

Though not the favored interpretation, photoionization models employed by \citet{Micheva20} and \citet{Dittenber20} do not rule out AGN contribution to the ionization field of Knot B. In a study of 15 LBAs and 40 Green Pea galaxies, \citet{Kim20} concluded that while intense central star formation may contribute heavily to galaxy-scale winds and feedback that can blow out channels for Ly$\alpha$ escape, the lack of significant correlations between star formation intensity and Ly$\alpha$ equivalent width or escape fraction hints that additional physical mechanisms may be necessary to fully explain LAEs. The lack of a detectable compact radio core at 2.3 GHz by \citet{Heisler98} places an upper bound on the mass of any putative AGN in the region. In \citet{Prestwich15}, we estimated an upper mass limit of $M_{\bullet} < 5\times10^{7} M_{\sun}$) for a black hole in X1 via the empirical relationship between the mass and the radio and X-ray fluxes \citep{Merloni05}. While the X-ray luminosities calculated for X1 above do show some variation from the original estimate, we note that the fundamental plane has a scatter of $\sim$1 dex, so our updated measurements are still consistent with the original mass estimate within the errors.

\subsection{The Nature of X-ray Sources in X1/Knot B}
    \label{sec:X1}

The best-fit photon index for the region of X1 is consistently hard. We note that our value for $\Gamma$ in observation 8175 is not as hard as was previously reported in \citet{Prestwich15} ($\Gamma$ = 1.2$\pm 0.2$); however, in this study we opted to increase the aperture size from r = 1.1\arcsec\ to 1.7\arcsec\ to model the region, so the spectra consequently capture more of the extended diffuse emission. \citet{BasuZych16} suggest that the intrinsic spectral slope of the point source(s) in X1 is therefore likely even flatter (harder) than what our fits give above. Such hard values for $\Gamma$ ($\sim$1.2) suggest that either one or both of the sources found in X1 could be a black hole binary in a low, hard state \citep{Remillard06, Sutton12}. In this state, a black hole would be expected to emit a compact jet, possibly detectable in spatially resolved radio observations. Given the low separation between X1a and X1b and the low count rates for the region, it is not feasible to characterize the sources individually via spectral analysis. However, if both X1a and X1b are LLAGN whose signatures are obscured by absorbing gas and dust, or masked by the intense star-formation, the two sources would constitute a rare local dual AGN system. At the estimated luminosities, they would be accreting in the radiatively inefficient accretion flow mode via advection \citep{Yuan14}, adding to the mechanical feedback in Knot B via winds. If instead the sources are a pair of closely separated XRBs, they might be "seed" BHs in the process of merging, leading to the formation of a SMBH. 

Considering the high luminosity of the X1a component during the first two observations ($L_{\rm X} \gtrsim 5 \times 10^{40}$ erg s$^{-1}$), we reiterate our previous suggestion that the region may contain an IMBH ($M_{\bullet} \gtrsim 7600 M_{\sun}$, for $\lambda_{\rm Edd} = 10\%$ (this represents a lower limit based on only the hard flux)). It is not possible to distinguish between an AGN or IMBH based on the spectra since the accretion mode of either does not depend on the black hole mass. In the low, hard accretion state, both scenarios would exhibit similar spectral characteristics. We can, however, conclude more definitively that the consistently hard spectrum is highly suggestive of sources in the low, hard state, as opposed to stellar-mass black holes accreting above the Eddington limit. Such accretion would likely exhibit a softer spectrum, unless preferentially viewed down the jet funnel to the hard central engine. Since none of the observations yield a soft spectrum for X1, we suggest the stellar-mass black hole case is a less likely explanation.

\subsection{The Nature of X-ray Sources in X2/ Knot C}
    \label{sec:X2}

The region of X2 also seems to contain 2 point sources. X2a dominates the hard flux at all times except for 16696 where X2b is comparable. The existence of two discrete sources also seems apparent in Figure \ref{fig:color_smooth}. X2b is undetected in the first two observations, but seems to increase in flux during a flare in the later two observations. Since X2a is brighter than X2b by a factor of $\geq$ 2 in most observations, we suggest that the spectral characteristics are mostly attributed to the dominant source, and we do not attempt to interpret X2b further. 

The best-fit photon index for the entire region of X2 is consistent with a value for the stacked spectrum ($\Gamma \sim$ 1.8) within the errors for all observations. Therefore, there does not seem to be spectral variability in X2.  

To further characterize the X2 region, we follow the scheme of \citet{Sazonov17} (and references therein) and compute the intrinsic soft/total X-ray flux ratios using our absorption-corrected flux estimates in Table \ref{tab:speclums}. We find values of 0.61$\pm 0.5$, 0.51$\pm 0.5$, 0.38$\pm 1.1$, and 0.62$\pm 0.5$ for the individual observations (in chronological order), and 0.53$\pm 0.2$ for the stacked spectrum. All observations are consistent within their errors with the value for the stacked spectrum.  Based on the demarcations from \citet{Sazonov17}, this categorizes the region of X2 as a soft ULX (0.6 $< F_{\rm soft}/F_{\rm broad} \leq 0.95$) during the first and last observations (although again, the error bars for {\it all} observations are consistent with this characterization). The best-fit photon index values for most observations are also consistent with $\Gamma \sim 2.1$ of the the empirically derived average X-ray spectrum of luminous HMXBs \citep{Sazonov17}.

X2a is by far the dominant source for three of the observations, in which it has $L_{\rm X} > 2 \times 10^{40}$ erg s$^{-1}$. In fact, during observation 16696 is the only time when the estimated $L_{\rm X}$ of X2a dips close to the threshold for soft ULXs \citep{Gladstone09}; so if X2a is a soft ULX, it would be confirmed to be a highly luminous one. This unusually high luminosity for a ULX might be evidence for similar, but more extreme class of accreting objects, Hyper-luminous X-ray sources \citep{Gao03}, whose luminosity can only be explained via super-Eddington accretion if a IMBH is invoked \citep{Swartz11, Sutton12, Kaaret17b}. This is consistent with the original picture in \citet{Prestwich15}: that the region of X2 might contain a lower range IMBH ($M_{\bullet} \gtrsim 20 M_{\sun}$). In this scenario, the softness is due to an optically thick Comptonized corona \citep{Gladstone09} which is formed as the inner region of the accretion disk is blown outwards by winds instigated by the super-Eddington accretion \citep{Middleton15}. The observed soft spectrum implies a line of site seen at least partially through these optically thick winds which Compton down-scatters the inner-disk photons (\citet{Sutton13}, see Figure 7 in \citet{Kaaret17b} for a schematic). We also note that in all observations, the spectrum for X2 exhibits a soft excess and a power law turnover around 3 keV, consistent with a ULX \citep{Gladstone09}. As well, the stacked spectrum and observation 8175 for X2 hint at evidence for soft line-like residuals due to partly ionized outflows expected for a ULX (see \citet{Kaaret17b} and references therein).

\citet{BasuZych16} conducted complex modelling of the X-ray Luminosity Function in Haro 11 based on observation 8175. \citet{BasuZych16} likewise suggest that X2 contains a ULX, but particularly that it is a Roche-Lobe Overflow high mass XRB (RLO HMXB, where the companion is a giant star) based on the age and metallicity of Knot C. As opposed to wind-fed HMXBs, RLO HMXBs can achieve stable accretion driven by the black hole at a mild ($\times$10) super-Eddington rate (\citet{BasuZych16} and references therein). \citet{Kaaret17b} estimate that the mean accretion rate at the Eddington limit ($\dot{m}_{\rm Edd} = 2.3 \times 10^{-8} M_{\bullet}M_{\sun}$yr$^{-1}$) can easily be exceeded by a 10$M_{\sun}$ black hole if the companion object is a massive star that evolves into the Hertzsprung Gap and expands to fill and thus overflow its Roche lobe. This seems to be the most likely explanation for X2a.
\citet{BasuZych13} suggest that HMXBs may account for the majority of the X-ray emission in Lyman break analogs, and that they are found preferentially in lower metallicity environments \citet{BasuZych16}. This corroborates the results of \citet{Brorby14}, who find that there is an enhancement in the HMXB population in blue compact dwarf galaxies (like Haro 11) over solar metallicity galaxies.

\section{Conclusions On the role of X-ray Sources in Lyman Leaking in Haro 11}
\label{sec:conclusions}

We have analyzed 3 new observations of the compact dwarf starburst galaxy Haro 11 and compared the results against our re-analysis of the original $Chandra$ observation. We once again find that the previously known X-ray sources X1 coincides with intense star-formation in Knot B and X2 coincides with Ly$\alpha$ emission in Knot C. Both X1 and X2 are best fit using a spatial model with 2 point sources and a low level of diffuse extended background. While our derived luminosities for the regions do not exhibit large-scale flux variability, the individual components do undergo noticeable periods of flaring and fading comparable to what is seen for similar sources in the starburst region of M82. In particular, the secondary component of X1 appears to brighten and surpass the primary component during the latter two observations. Meanwhile, the soft primary component of X2 fades by 80\% during observation 16696, allowing the secondary source to shine through. 

We conclude by posing the question: are X-ray sources related to Ly$\alpha$ and continuum escape? Both regions X1 and X2 are in the midst of highly active star-formation that should yield UV bright stars capable of producing Lyman continuum and line emission. Both likely contain XRBs that contribute to the mechanical feedback of their respective regions; however, while the ages of the current star-forming episodes are similar ( $\sim$3.5$-$4.3 Myr in Knot B vs. $\sim$7.8$-$10 Myr in Knot C \citep{James13, Adamo10}), the rate of current star-formation is vastly different (0.86 vs. 0.09 $M_{\sun}$ yr$^{-1}$) \citep{James13}. One might then expect the more active star formation in Knot B to lead to stronger winds from supernovae and high mass stars, and thus a higher chance of Ly$\alpha$ escape. But no strong Ly$\alpha$ emission is observed coming from Knot B. However, in Knot C, the Ly$\alpha$ escape fraction is found to be $\sim$3\% \citep{Hayes07}. 

Based on the spectral and spatial analysis of Haro 11, X2 most likely contains a soft ULX super-Eddington source capable of powerful winds the mechanical power to blow out substantial material from the surrounding region. While X1 might contain LLAGN, the evidence is ambiguous. If both sources in X1 are XRBs in the low, hard state, then they would not be producing the winds necessary for a blow-out scenario.  

We refer the reader to our previous argument in \citet{Prestwich15} (see Figure 8 therein) which puts X2 in the context of mechanical power using the Starburst99 code \citep{Leitherer10}. The relatively high luminosities of X2a and b both fall above the estimated threshold where XRB winds dominate over supernovae and stellar winds in the overall mechanical luminosity budget for the starburst region of Knot C. Specifically, our estimates for the luminosity of X2a are at or above $10^{40}$ erg s$^{-1}$ for all observations. Assuming that the accreting objects in X2 are producing outflows with mechanical power at least equal to their radiative luminosities \citep{Gallo05, Justham12}, the feedback should be comparable to, if not dominant over, the feedback from star-formation given the current age of the cluster in Knot C \citep{Adamo10, James13}. We therefore reiterate our previous conclusion that mechanical luminosity from ULXs may be a key mechanism to aid in the blow-out of neutral Hydrogen medium, allowing the escape of the Ly$\alpha$ emission observed in the X2 region. 

A similar conclusion about the importance of XRB feedback in facilitating Lyman emission was reached by \citet{Bluem19}. In their sample of 8 blue compact galaxies studied as potential Lyman continuum emitters, they suggest the importance of XRBs can be assessed based on an excess of X-ray flux above the amount predicted from star formation via eq. 22 from \citet{Mineo12}. They find that the $L_{\rm X}$ for their galaxies is above this $L_{\rm XRB} - $SFR relation, and suggest that there might be an excess of XRBs in the regions which are causing the mechanical feedback and thus LyC/Ly$\alpha$ leackage. Using the SFR for Haro 11 Knot C calculated by \citet{James13} we find an expected $L_{\rm 0.5-8 keV}^{\rm XRB} = 2.35 \times 10^{38} $erg s$^{-1}$, which is $\lesssim$ 2 dex lower than the broad band luminosities found for any observation of X2, given in Table \ref{tab:speclums}. Given the compact nature of Knot C paired with the the softness of X2, we suggest that the bulk of this excess X-ray flux can be explained by an individual ultra-luminous source, X2a, with additional flux from X2b. A potential caveat to this is that excess $L_{\rm X}$ could be attributed to an LLAGN (see \citet{Gross19} for a similar argument). X2 as a whole does fall within the AGN/composite region of Figure \ref{fig:BPT}, but this may be due to contribution from source X2b, and not the soft ULX X2a which we suggest is the major source of mechanical feedback via winds and subsequent Ly$\alpha$ escape in X2. For an example of a recently discovered obscured AGN at $z \sim 4$ that {\it has} been linked to Ly$\alpha$ emission, see \citet{Vito20}.

The consequences of these different nuclear engines can also be seen in the ionization of Knot C. \citet{RiveraThorsen17} find that it is a highly ionized, density-bounded, low column density region, which likely makes Ly$\alpha$ escape more accessible \citep{Micheva20}. There is even evidence of a lowly-ionized, large scale, fragmented superbubble surrounding the entirety of Knot C \citep{Menacho19}. A similar conclusion has been suggested for the low-metallicity starburst galaxy ESO 338-4, which might harbor a ULX powered by an IMBH of $M_{\bullet} \gtrsim 300 M_{\sun}$ \citep{Oskinova19}. Meanwhile, Haro 11 Knot B has the highest extinction of all three knot regions \citep{Adamo10}, which would at least absorb some portion of Lyman line and continuum emission. This implies that powerful winds from the ULX in X2 are the driver of Ly$\alpha$ escape in Haro 11. We therefore suggest that ULXs may be a necessary ingredient in other LAEs, and analogously higher redshift Lyman break galaxies.

Lyman continuum escape is also observed from Haro 11, with an escape fraction of $\sim$3.3$-$9\% \citep{Leitet11, Hayes07}, but isolating which starforming knot may be the source has yet proved elusive. Recent ionization parameter mapping of Haro 11 by \citet{Keenan17} suggests that the starburst region of Knot A might be responsible for the Lyman Continuum escape. While they do not find compelling evidence of LyC in Knots B or C, they suggest that Knot A contains optically thin ionized regions that would facilitate LyC escape. We note however, that more recent studies of LyC emitting galaxies indicate that the ionization parameter [\ion{O}{iii}]/[\ion{O}{ii}] is not correlated with LyC escape fraction (\citet{Bassett19} and references therein), which casts doubt on the interpretation of Knot A being the origin LyC emission. \citet{Bik15} also find evidence for extremely ionized regions concentrated in the Southwest of Haro 11 around Knot A, due to density-bounded ionized bubbles around individual super star clusters. If Knot A is conclusively shown to be the source of LyC leakage, its separation from Knot C implies that the observed LyC and Ly$\alpha$ emissions are at least partially decoupled as Ly$\alpha$ is gradually diffused through many resonant scatterings \citep{Hayes07}. 

Given the younger estimated age of the starburst in Knot A ($\sim$ 1$-$4.9 Myr \citep{Keenan17, James13}), it is unsurprising that we do not detect any coincident X-ray point sources (and thus no ULXs) in our analysis. As noted by \citet{Kaaret17}, the requisite time for the evolution of a supergiant and thus the formation of a HMXB is much too long ($\sim 12$Myr) to match the young age of the starburst in Knot A. Even in the compact star cluster of MGG 11 ($r \sim\ 1.2 $pc), simulations of dynamical friction-induced stellar collisions did not result in an IMBH within the first $\sim$3 Myr \citep{PortegiesZwart04}.

One of the only other local examples of a Lyman continuum emitter is Tol 1247-232 \citep{Kaaret17}. Interestingly, Tol 1247 has a hard spectrum like X1a and b, as opposed to the soft luminous ULX in X2a. The hard $\Gamma$ also reinforces the picture of an AGN or IMBH which is being viewed "down the barrel" of the jet, without the softer X-ray component due to scattering seen at higher inclination angles. We show it's placement on Figure \ref{fig:BPT} based on \citet{Leitet13}. Tol 1247 has emission line characteristics of a purely star-forming region, although the very young stars in the central region of the galaxy ($\sim$3 Myr, similar to Haro 11 Knot B) likewise suggest that a LLAGN signature might be masked by starburst effects.\ \citet{Kaaret17} suggest that a combination of X-ray point sources (LLAGN, XRBs, etc.) and intense mechanical feedback from star-formation may be responsible for the LyC leakage in Tol 1247. As seen in Haro 11, \citet{Micheva18} suggest that two-stage starbursts may be a common feature of LyC emitters, finding that Tol 1247 also contains separate generations of star clusters which have been responsible for sustaining optically thin channels for LyC escape at different times. As well, a two-stage starburst may allow enough the time to initially grow a ULX in the first starburst phase, which then has the mechanical power to blow away material to allow the escape of UV photons produced during the second starburst phase.  


Given the significance of Haro 11 as the best local example of a LAE and Lyman continuum emitter, definitive identification of AGN in both Knots B and C would be useful in further understanding the mechanics at play. Specifically, high spatial resolution radio interferometry observations could give conclusive evidence of compact steep spectrum cores associated with any LLAGN or IMBH (as was done in \citet{Alexandroff12}), or jets providing feedback. It could also potentially reveal the spatial distribution of multiple accreting sources in X1, if there are indeed more than one as our analysis suggests. 
Additionally, the NIRSPEC instrument on board the $James\ Webb\ Space\ Telescope$ could be useful to obtain further IFU spectra of each Knot. The spatial resolution of 0.1\arcsec\ would be able to discern small-scale emission features in the 0.6 to 5.0 $\mu$m range, giving greater insight as to differences in the complex ionized structures in Knots B and C, and their links to aiding Ly$\alpha$ escape.

\section*{Acknowledgements}

The authors thank the Chandra X-ray Center Help Desk for insightful conversations on data analysis. The authors also thank the anonymous reviewer whose comments improved the manuscript. 
Support for this work was provided by the National Aeronautics and
Space Administration through Chandra Award Number GO5-16028B issued by the
Chandra X-ray Observatory Center, which is operated by the Smithsonian
Astrophysical Observatory for and on behalf of the National
Aeronautics Space Administration under contract NAS8-03060.

\section*{Data Availability}

Data are available upon reasonable request to the corresponding author.



\bibliographystyle{mnras}
\bibliography{example} 

\begin{thebibliography}{}
\makeatletter
\relax
\def\mn@urlcharsother{\let\do\@makeother \do\$\do\&\do\#\do\^\do\_\do\%\do\~}
\def\mn@doi{\begingroup\mn@urlcharsother \@ifnextchar [ {\mn@doi@}
  {\mn@doi@[]}}
\def\mn@doi@[#1]#2{\def\@tempa{#1}\ifx\@tempa\@empty \href
  {http://dx.doi.org/#2} {doi:#2}\else \href {http://dx.doi.org/#2} {#1}\fi
  \endgroup}
\def\mn@eprint#1#2{\mn@eprint@#1:#2::\@nil}
\def\mn@eprint@arXiv#1{\href {http://arxiv.org/abs/#1} {{\tt arXiv:#1}}}
\def\mn@eprint@dblp#1{\href {http://dblp.uni-trier.de/rec/bibtex/#1.xml}
  {dblp:#1}}
\def\mn@eprint@#1:#2:#3:#4\@nil{\def\@tempa {#1}\def\@tempb {#2}\def\@tempc
  {#3}\ifx \@tempc \@empty \let \@tempc \@tempb \let \@tempb \@tempa \fi \ifx
  \@tempb \@empty \def\@tempb {arXiv}\fi \@ifundefined
  {mn@eprint@\@tempb}{\@tempb:\@tempc}{\expandafter \expandafter \csname
  mn@eprint@\@tempb\endcsname \expandafter{\@tempc}}}

\bibitem[\protect\citeauthoryear{{Abbott} et~al.,}{{Abbott}
  et~al.}{2020a}]{Abbott20a}
{Abbott} R.,  et~al., 2020a, \prl, 125, 101102

\bibitem[\protect\citeauthoryear{{Abbott} et~al.,}{{Abbott}
  et~al.}{2020b}]{Abbott20b}
{Abbott} R.,  et~al., 2020b, \apjl, 900, L13

\bibitem[\protect\citeauthoryear{Abolmasov}{Abolmasov}{2011}]{Abolmasov11}
Abolmasov P.,  2011, New Astronomy, 16, 138

\bibitem[\protect\citeauthoryear{Abolmasov, Swartz, Fabrika, Ghosh, Sholukhova
  \& Tennant}{Abolmasov et~al.}{2007}]{Abolmasov07}
Abolmasov P.~K.,  Swartz D.~A.,  Fabrika S.,  Ghosh K.~K.,  Sholukhova O.,
  Tennant A.~F.,  2007, The Astrophysical Journal, 668, 124

\bibitem[\protect\citeauthoryear{Adamo, {\"O}stlin, Zackrisson, Hayes, Cumming
  \& Micheva}{Adamo et~al.}{2010}]{Adamo10}
Adamo A.,  {\"O}stlin G.,  Zackrisson E.,  Hayes M.,  Cumming R.~J.,   Micheva
  G.,  2010, Monthly Notices of the Royal Astronomical Society, 407, 870

\bibitem[\protect\citeauthoryear{Alavi et~al.,}{Alavi et~al.}{2016}]{Alavi16}
Alavi A.,  et~al., 2016, The Astrophysical Journal, 832, 56

\bibitem[\protect\citeauthoryear{Alexandroff et~al.,}{Alexandroff
  et~al.}{2012}]{Alexandroff12}
Alexandroff R.,  et~al., 2012, Monthly Notices of the Royal Astronomical
  Society, 423, 1325

\bibitem[\protect\citeauthoryear{{Baldwin}, {Phillips}  \&
  {Terlevich}}{{Baldwin} et~al.}{1981}]{Baldwin81}
{Baldwin} J.~A.,  {Phillips} M.~M.,   {Terlevich} R.,  1981, \pasp, 93, 5

\bibitem[\protect\citeauthoryear{{Bassett} et~al.,}{{Bassett}
  et~al.}{2019}]{Bassett19}
{Bassett} R.,  et~al., 2019, \mnras, 483, 5223

\bibitem[\protect\citeauthoryear{Basu-Zych et~al.,}{Basu-Zych
  et~al.}{2013}]{BasuZych13}
Basu-Zych A.~R.,  et~al., 2013, The Astrophysical Journal, 774, 152

\bibitem[\protect\citeauthoryear{Basu-Zych, Lehmer, Fragos, Hornschemeier,
  Yukita, Zezas  \& Ptak}{Basu-Zych et~al.}{2016}]{BasuZych16}
Basu-Zych A.~R.,  Lehmer B.,  Fragos T.,  Hornschemeier A.,  Yukita M.,  Zezas
  A.,   Ptak A.,  2016, The Astrophysical Journal, 818, 140

\bibitem[\protect\citeauthoryear{{Bergvall}, {Masegosa}, {{\"O}stlin}  \&
  {Cernicharo}}{{Bergvall} et~al.}{2000}]{Bergvall00}
{Bergvall} N.,  {Masegosa} J.,  {{\"O}stlin} G.,   {Cernicharo} J.,  2000,
  \aap, 359, 41

\bibitem[\protect\citeauthoryear{{Bergvall}, {Zackrisson}, {Andersson},
  {Arnberg}, {Masegosa}  \& {{\"O}stlin}}{{Bergvall} et~al.}{2006}]{Bergvall06}
{Bergvall} N.,  {Zackrisson} E.,  {Andersson} B.~G.,  {Arnberg} D.,  {Masegosa}
  J.,   {{\"O}stlin} G.,  2006, \aap, 448, 513

\bibitem[\protect\citeauthoryear{Bian, Kewley, Groves  \& Dopita}{Bian
  et~al.}{2020}]{Bian20}
Bian F.,  Kewley L.~J.,  Groves B.,   Dopita M.~A.,  2020, Monthly Notices of
  the Royal Astronomical Society, 493, 580

\bibitem[\protect\citeauthoryear{{Bik}, {{\"O}stlin}, {Menacho}, {Adamo},
  {Hayes}, {Melinder}  \& {Amram}}{{Bik} et~al.}{2015}]{Bik15}
{Bik} A.,  {{\"O}stlin} G.,  {Menacho} V.,  {Adamo} A.,  {Hayes} M.,
  {Melinder} J.,   {Amram} P.,  2015, arXiv e-prints, p. arXiv:1510.01944

\bibitem[\protect\citeauthoryear{Bluem, Kaaret, Prestwich  \& Brorby}{Bluem
  et~al.}{2019}]{Bluem19}
Bluem J.,  Kaaret P.,  Prestwich A.,   Brorby M.,  2019, Monthly Notices of the
  Royal Astronomical Society, 487, 4093

\bibitem[\protect\citeauthoryear{Bosman, Fan, Jiang, Reed, Matsuoka, Becker  \&
  Haehnelt}{Bosman et~al.}{2018}]{Bosman18}
Bosman S. E.~I.,  Fan X.,  Jiang L.,  Reed S.,  Matsuoka Y.,  Becker G.,
  Haehnelt M.,  2018, Monthly Notices of the Royal Astronomical Society, 479,
  1055

\bibitem[\protect\citeauthoryear{Brorby, Kaaret  \& Prestwich}{Brorby
  et~al.}{2014}]{Brorby14}
Brorby M.,  Kaaret P.,   Prestwich A.,  2014, Monthly Notices of the Royal
  Astronomical Society, 441, 2346

\bibitem[\protect\citeauthoryear{{Carter}, {Karovska}, {Jerius}, {Glotfelty}
  \& {Beikman}}{{Carter} et~al.}{2003}]{Carter03}
{Carter} C.,  {Karovska} M.,  {Jerius} D.,  {Glotfelty} K.,   {Beikman} S.,
  2003, in {Payne} H.~E.,  {Jedrzejewski} R.~I.,   {Hook} R.~N.,  eds,
  Astronomical Society of the Pacific Conference Series Vol. 295, Astronomical
  Data Analysis Software and Systems XII. p.~477

\bibitem[\protect\citeauthoryear{{Cash}}{{Cash}}{1979}]{Cash79}
{Cash} W.,  1979, \apj, 228, 939

\bibitem[\protect\citeauthoryear{{Chiang} \& {Kong}}{{Chiang} \&
  {Kong}}{2011}]{Chiang11}
{Chiang} Y.-K.,  {Kong} A. K.~H.,  2011, \mnras, 414, 1329

\bibitem[\protect\citeauthoryear{{Chilingarian}, {Katkov}, {Zolotukhin},
  {Grishin}, {Beletsky}, {Boutsia}  \& {Osip}}{{Chilingarian}
  et~al.}{2018}]{Chilingarian18}
{Chilingarian} I.~V.,  {Katkov} I.~Y.,  {Zolotukhin} I.~Y.,  {Grishin} K.~A.,
  {Beletsky} Y.,  {Boutsia} K.,   {Osip} D.~J.,  2018, \apj, 863, 1

\bibitem[\protect\citeauthoryear{{Davis} et~al.,}{{Davis}
  et~al.}{2012}]{Davis12}
{Davis} J.~E.,  et~al., 2012, in {Takahashi} T.,  {Murray} S.~S.,   {den
  Herder} J.-W.~A.,  eds,  Society of Photo-Optical Instrumentation Engineers
  (SPIE) Conference Series Vol. 8443, Space Telescopes and Instrumentation
  2012: Ultraviolet to Gamma Ray. p. 84431A

\bibitem[\protect\citeauthoryear{{De Rossi}, {Rieke}, {Shivaei}, {Bromm}  \&
  {Lyu}}{{De Rossi} et~al.}{2018}]{DeRossi18}
{De Rossi} M.~E.,  {Rieke} G.~H.,  {Shivaei} I.,  {Bromm} V.,   {Lyu} J.,
  2018, \apj, 869, 4

\bibitem[\protect\citeauthoryear{Dickey \& Lockman}{Dickey \&
  Lockman}{1990}]{Dickey90}
Dickey J.~M.,  Lockman F.~J.,  1990, Annual Review of Astronomy and
  Astrophysics, 28, 215

\bibitem[\protect\citeauthoryear{Dittenber, Oey, Hodges-Kluck, Gallo, Hayes,
  {\"O}stlin  \& Melinder}{Dittenber et~al.}{2020}]{Dittenber20}
Dittenber B.,  Oey M.~S.,  Hodges-Kluck E.,  Gallo E.,  Hayes M.,  {\"O}stlin
  G.,   Melinder J.,  2020, The Astrophysical Journal, 890, L12

\bibitem[\protect\citeauthoryear{Dunkley et~al.,}{Dunkley
  et~al.}{2009}]{Dunkley09}
Dunkley J.,  et~al., 2009, The Astrophysical Journal Supplement Series, 180,
  306

\bibitem[\protect\citeauthoryear{Erb, Pettini, Steidel, Strom, Rudie, Trainor,
  Shapley  \& Reddy}{Erb et~al.}{2016}]{Erb16}
Erb D.~K.,  Pettini M.,  Steidel C.~C.,  Strom A.~L.,  Rudie G.~C.,  Trainor
  R.~F.,  Shapley A.~E.,   Reddy N.~A.,  2016, The Astrophysical Journal, 830,
  52

\bibitem[\protect\citeauthoryear{{Freedman} et~al.,}{{Freedman}
  et~al.}{1994}]{Freedman94}
{Freedman} W.~L.,  et~al., 1994, \apj, 427, 628

\bibitem[\protect\citeauthoryear{{Gallagher} \& {Smith}}{{Gallagher} \&
  {Smith}}{1999}]{Gallagher99}
{Gallagher} John~S. I.,  {Smith} L.~J.,  1999, \mnras, 304, 540

\bibitem[\protect\citeauthoryear{{Gallo}, {Fender}, {Kaiser}, {Russell},
  {Morganti}, {Oosterloo}  \& {Heinz}}{{Gallo} et~al.}{2005}]{Gallo05}
{Gallo} E.,  {Fender} R.,  {Kaiser} C.,  {Russell} D.,  {Morganti} R.,
  {Oosterloo} T.,   {Heinz} S.,  2005, \nat, 436, 819

\bibitem[\protect\citeauthoryear{{Gao}, {Wang}, {Appleton}  \& {Lucas}}{{Gao}
  et~al.}{2003}]{Gao03}
{Gao} Y.,  {Wang} Q.~D.,  {Appleton} P.~N.,   {Lucas} R.~A.,  2003, \apjl, 596,
  L171

\bibitem[\protect\citeauthoryear{Gladstone, Roberts  \& Done}{Gladstone
  et~al.}{2009}]{Gladstone09}
Gladstone J.~C.,  Roberts T.~P.,   Done C.,  2009, Monthly Notices of the Royal
  Astronomical Society, 397, 1836

\bibitem[\protect\citeauthoryear{{Grimes} et~al.,}{{Grimes}
  et~al.}{2007}]{Grimes07}
{Grimes} J.~P.,  et~al., 2007, \apj, 668, 891

\bibitem[\protect\citeauthoryear{Gross, Fu, Myers, Wrobel  \& Djorgovski}{Gross
  et~al.}{2019}]{Gross19}
Gross A.~C.,  Fu H.,  Myers A.~D.,  Wrobel J.~M.,   Djorgovski S.~G.,  2019,
  The Astrophysical Journal, 883, 50

\bibitem[\protect\citeauthoryear{{Guseva}, {Izotov}, {Fricke}  \&
  {Henkel}}{{Guseva} et~al.}{2012}]{Guseva12}
{Guseva} N.~G.,  {Izotov} Y.~I.,  {Fricke} K.~J.,   {Henkel} C.,  2012, \aap,
  541, A115

\bibitem[\protect\citeauthoryear{Guseva et~al.,}{Guseva
  et~al.}{2020}]{Guseva20}
Guseva N.~G.,  et~al., 2020, Monthly Notices of the Royal Astronomical Society,
  497, 4293

\bibitem[\protect\citeauthoryear{{Hayes}, {{\"O}stlin}, {Atek}, {Kunth},
  {Mas-Hesse}, {Leitherer}, {Jim{\'e}nez-Bail{\'o}n}  \& {Adamo}}{{Hayes}
  et~al.}{2007}]{Hayes07}
{Hayes} M.,  {{\"O}stlin} G.,  {Atek} H.,  {Kunth} D.,  {Mas-Hesse} J.~M.,
  {Leitherer} C.,  {Jim{\'e}nez-Bail{\'o}n} E.,   {Adamo} A.,  2007, \mnras,
  382, 1465

\bibitem[\protect\citeauthoryear{Hayes et~al.,}{Hayes et~al.}{2010}]{Hayes10}
Hayes M.,  et~al., 2010, Nature, 464, 562

\bibitem[\protect\citeauthoryear{{Heckman} et~al.,}{{Heckman}
  et~al.}{2005}]{Heckman05}
{Heckman} T.~M.,  et~al., 2005, \apjl, 619, L35

\bibitem[\protect\citeauthoryear{Heckman et~al.,}{Heckman
  et~al.}{2011}]{Heckman11}
Heckman T.~M.,  et~al., 2011, The Astrophysical Journal, 730, 5

\bibitem[\protect\citeauthoryear{{Heisler}, {Norris}, {Jauncey}, {Reynolds}  \&
  {King}}{{Heisler} et~al.}{1998}]{Heisler98}
{Heisler} C.~A.,  {Norris} R.~P.,  {Jauncey} D.~L.,  {Reynolds} J.~E.,   {King}
  E.~A.,  1998, \mnras, 300, 1111

\bibitem[\protect\citeauthoryear{{Ho}}{{Ho}}{2008}]{Ho08}
{Ho} L.~C.,  2008, \araa, 46, 475

\bibitem[\protect\citeauthoryear{{Hoopes} et~al.,}{{Hoopes}
  et~al.}{2007}]{Hoopes07}
{Hoopes} C.~G.,  et~al., 2007, \apjs, 173, 441

\bibitem[\protect\citeauthoryear{{Izotov}, {Guseva}, {Fricke}, {Henkel},
  {Schaerer}  \& {Thuan}}{{Izotov} et~al.}{2021}]{Izotov21}
{Izotov} Y.~I.,  {Guseva} N.~G.,  {Fricke} K.~J.,  {Henkel} C.,  {Schaerer} D.,
    {Thuan} T.~X.,  2021, \aap, 646, A138

\bibitem[\protect\citeauthoryear{James, Tsamis, Walsh, Barlow  \&
  Westmoquette}{James et~al.}{2013}]{James13}
James B.~L.,  Tsamis Y.~G.,  Walsh J.~R.,  Barlow M.~J.,   Westmoquette M.~S.,
  2013, Monthly Notices of the Royal Astronomical Society, 430, 2097

\bibitem[\protect\citeauthoryear{Jia, Ptak, Heckman, Overzier, Hornschemeier
  \& LaMassa}{Jia et~al.}{2011}]{Jia11}
Jia J.,  Ptak A.,  Heckman T.~M.,  Overzier R.~A.,  Hornschemeier A.,   LaMassa
  S.~M.,  2011, The Astrophysical Journal, 731, 55

\bibitem[\protect\citeauthoryear{Justham \& Schawinski}{Justham \&
  Schawinski}{2012}]{Justham12}
Justham S.,  Schawinski K.,  2012, Monthly Notices of the Royal Astronomical
  Society, 423, 1641

\bibitem[\protect\citeauthoryear{{Kaaret}, {Prestwich}, {Zezas}, {Murray},
  {Kim}, {Kilgard}, {Schlegel}  \& {Ward}}{{Kaaret} et~al.}{2001}]{Kaaret01}
{Kaaret} P.,  {Prestwich} A.~H.,  {Zezas} A.,  {Murray} S.~S.,  {Kim} D.~W.,
  {Kilgard} R.~E.,  {Schlegel} E.~M.,   {Ward} M.~J.,  2001, \mnras, 321, L29

\bibitem[\protect\citeauthoryear{Kaaret, Feng  \& Gorski}{Kaaret
  et~al.}{2009}]{Kaaret09}
Kaaret P.,  Feng H.,   Gorski M.,  2009, The Astrophysical Journal, 692, 653

\bibitem[\protect\citeauthoryear{Kaaret, Feng  \& Roberts}{Kaaret
  et~al.}{2017a}]{Kaaret17b}
Kaaret P.,  Feng H.,   Roberts T.~P.,  2017a, Annual Review of Astronomy and
  Astrophysics, 55, 303

\bibitem[\protect\citeauthoryear{Kaaret, Brorby, Casella  \& Prestwich}{Kaaret
  et~al.}{2017b}]{Kaaret17}
Kaaret P.,  Brorby M.,  Casella L.,   Prestwich A.~H.,  2017b, Monthly Notices
  of the Royal Astronomical Society, 471, 4234

\bibitem[\protect\citeauthoryear{{Kauffmann} et~al.,}{{Kauffmann}
  et~al.}{2003}]{Kauffmann03}
{Kauffmann} G.,  et~al., 2003, \mnras, 346, 1055

\bibitem[\protect\citeauthoryear{Keenan, Oey, Jaskot  \& James}{Keenan
  et~al.}{2017}]{Keenan17}
Keenan R.~P.,  Oey M.~S.,  Jaskot A.~E.,   James B.~L.,  2017, The
  Astrophysical Journal, 848, 12

\bibitem[\protect\citeauthoryear{{Kewley}, {Dopita}, {Sutherland}, {Heisler}
  \& {Trevena}}{{Kewley} et~al.}{2001}]{Kewley01}
{Kewley} L.~J.,  {Dopita} M.~A.,  {Sutherland} R.~S.,  {Heisler} C.~A.,
  {Trevena} J.,  2001, \apj, 556, 121

\bibitem[\protect\citeauthoryear{{Kim}, {Malhotra}, {Rhoads}, {Butler}  \&
  {Yang}}{{Kim} et~al.}{2020}]{Kim20}
{Kim} K.,  {Malhotra} S.,  {Rhoads} J.~E.,  {Butler} N.~R.,   {Yang} H.,  2020,
  \apj, 893, 134

\bibitem[\protect\citeauthoryear{Kunth, Leitherer, Mas-Hesse, Ostlin  \&
  Petrosian}{Kunth et~al.}{2003}]{Kunth03}
Kunth D.,  Leitherer C.,  Mas-Hesse J.~M.,  Ostlin G.,   Petrosian A.,  2003,
  The Astrophysical Journal, 597, 263

\bibitem[\protect\citeauthoryear{{Leitet, E.}, {Bergvall, N.}, {Piskunov, N.}
  \& {Andersson, B.-G.}}{{Leitet, E.} et~al.}{2011}]{Leitet11}
{Leitet, E.} {Bergvall, N.} {Piskunov, N.}  {Andersson, B.-G.} 2011, A\&A, 532,
  A107

\bibitem[\protect\citeauthoryear{{Leitet, E.}, {Bergvall, N.}, {Hayes, M.},
  {Linn\'e, S.}  \& {Zackrisson, E.}}{{Leitet, E.} et~al.}{2013}]{Leitet13}
{Leitet, E.} {Bergvall, N.} {Hayes, M.} {Linn\'e, S.}  {Zackrisson, E.} 2013,
  A\&A, 553, A106

\bibitem[\protect\citeauthoryear{Leitherer, Ot{\'{a}}lvaro, Bresolin,
  Kudritzki, Faro, Pauldrach, Pettini  \& Rix}{Leitherer
  et~al.}{2010}]{Leitherer10}
Leitherer C.,  Ot{\'{a}}lvaro P. A.~O.,  Bresolin F.,  Kudritzki R.-P.,  Faro
  B.~L.,  Pauldrach A. W.~A.,  Pettini M.,   Rix S.~A.,  2010, The
  Astrophysical Journal Supplement Series, 189, 309

\bibitem[\protect\citeauthoryear{Leitherer, Hernandez, Lee  \& Oey}{Leitherer
  et~al.}{2016}]{Leitherer16}
Leitherer C.,  Hernandez S.,  Lee J.~C.,   Oey M.~S.,  2016, The Astrophysical
  Journal, 823, 64

\bibitem[\protect\citeauthoryear{{Li}, {Kastner}, {Prigozhin}, {Schulz},
  {Feigelson}  \& {Getman}}{{Li} et~al.}{2004}]{Li04}
{Li} J.,  {Kastner} J.~H.,  {Prigozhin} G.~Y.,  {Schulz} N.~S.,  {Feigelson}
  E.~D.,   {Getman} K.~V.,  2004, \apj, 610, 1204

\bibitem[\protect\citeauthoryear{Loaiza-Agudelo, Overzier  \&
  Heckman}{Loaiza-Agudelo et~al.}{2020}]{LoaizaAgudelo20}
Loaiza-Agudelo M.,  Overzier R.~A.,   Heckman T.~M.,  2020, The Astrophysical
  Journal, 891, 19

\bibitem[\protect\citeauthoryear{Loeb}{Loeb}{2010}]{Loeb10}
Loeb A.,  2010, How Did the First Stars and Galaxies Form?.
Princeton Univ. Press, Princeton, NJ

\bibitem[\protect\citeauthoryear{{Makishima} et~al.,}{{Makishima}
  et~al.}{2000}]{Makishima00}
{Makishima} K.,  et~al., 2000, \apj, 535, 632

\bibitem[\protect\citeauthoryear{{Mallery} et~al.,}{{Mallery}
  et~al.}{2012}]{Mallery12}
{Mallery} R.~P.,  et~al., 2012, \apj, 760, 128

\bibitem[\protect\citeauthoryear{{Mas-Hesse}, {Kunth}, {Tenorio-Tagle},
  {Leitherer}, {Terlevich}  \& {Terlevich}}{{Mas-Hesse}
  et~al.}{2003}]{MasHesse03}
{Mas-Hesse} J.~M.,  {Kunth} D.,  {Tenorio-Tagle} G.,  {Leitherer} C.,
  {Terlevich} R.~J.,   {Terlevich} E.,  2003, \apj, 598, 858

\bibitem[\protect\citeauthoryear{Matsumoto, Tsuru, Koyama, Awaki, Canizares,
  Kawai, Matsushita  \& Kawabe}{Matsumoto et~al.}{2001}]{Matsumoto01}
Matsumoto H.,  Tsuru T.~G.,  Koyama K.,  Awaki H.,  Canizares C.~R.,  Kawai N.,
   Matsushita S.,   Kawabe R.,  2001, The Astrophysical Journal, 547, L25

\bibitem[\protect\citeauthoryear{Menacho et~al.,}{Menacho
  et~al.}{2019}]{Menacho19}
Menacho V.,  et~al., 2019, Monthly Notices of the Royal Astronomical Society,
  487, 3183

\bibitem[\protect\citeauthoryear{Merloni, Heinz  \& Matteo}{Merloni
  et~al.}{2005}]{Merloni05}
Merloni A.,  Heinz S.,   Matteo T.~D.,  2005, Astrophysics and Space Science,
  300, 45

\bibitem[\protect\citeauthoryear{Mesinger, Ferrara  \& Spiegel}{Mesinger
  et~al.}{2013}]{Mesinger13}
Mesinger A.,  Ferrara A.,   Spiegel D.~S.,  2013, Monthly Notices of the Royal
  Astronomical Society, 431, 621

\bibitem[\protect\citeauthoryear{{Mezcua} \& {Dom{\'\i}nguez
  S{\'a}nchez}}{{Mezcua} \& {Dom{\'\i}nguez S{\'a}nchez}}{2020}]{Mezcua20}
{Mezcua} M.,  {Dom{\'\i}nguez S{\'a}nchez} H.,  2020, \apjl, 898, L30

\bibitem[\protect\citeauthoryear{Micheva, Oey, Keenan, Jaskot  \&
  James}{Micheva et~al.}{2018}]{Micheva18}
Micheva G.,  Oey M.~S.,  Keenan R.~P.,  Jaskot A.~E.,   James B.~L.,  2018, The
  Astrophysical Journal, 867, 2

\bibitem[\protect\citeauthoryear{{Micheva} et~al.,}{{Micheva}
  et~al.}{2020}]{Micheva20}
{Micheva} G.,  et~al., 2020, \apj, 903, 123

\bibitem[\protect\citeauthoryear{Middleton, Heil, Pintore, Walton  \&
  Roberts}{Middleton et~al.}{2015}]{Middleton15}
Middleton M.~J.,  Heil L.,  Pintore F.,  Walton D.~J.,   Roberts T.~P.,  2015,
  Monthly Notices of the Royal Astronomical Society, 447, 3243

\bibitem[\protect\citeauthoryear{{Mineo}, {Gilfanov}  \& {Sunyaev}}{{Mineo}
  et~al.}{2012}]{Mineo12}
{Mineo} S.,  {Gilfanov} M.,   {Sunyaev} R.,  2012, \mnras, 419, 2095

\bibitem[\protect\citeauthoryear{{Morrison} \& {McCammon}}{{Morrison} \&
  {McCammon}}{1983}]{Morrison83}
{Morrison} R.,  {McCammon} D.,  1983, \apj, 270, 119

\bibitem[\protect\citeauthoryear{{Neufeld}}{{Neufeld}}{1991}]{Neufeld91}
{Neufeld} D.~A.,  1991, \apjl, 370, L85

\bibitem[\protect\citeauthoryear{{Nilsson, K. K.} et~al.,}{{Nilsson, K. K.}
  et~al.}{2007}]{Nilsson07}
{Nilsson, K. K.} et~al., 2007, A\&A, 471, 71

\bibitem[\protect\citeauthoryear{Orsi, Lacey  \& Baugh}{Orsi
  et~al.}{2012}]{Orsi12}
Orsi A.,  Lacey C.~G.,   Baugh C.~M.,  2012, Monthly Notices of the Royal
  Astronomical Society, 425, 87

\bibitem[\protect\citeauthoryear{{Oskinova, L. M.}, {Bik, A.}, {Mas-Hesse, J.
  M.}, {Hayes, M.}, {Adamo, A.}, {\"Ostlin, G.}, {F\"urst, F.}  \&
  {Ot\'{\i}-Floranes, H.}}{{Oskinova, L. M.} et~al.}{2019}]{Oskinova19}
{Oskinova, L. M.} {Bik, A.} {Mas-Hesse, J. M.} {Hayes, M.} {Adamo, A.}
  {\"Ostlin, G.} {F\"urst, F.}  {Ot\'{\i}-Floranes, H.} 2019, A\&A, 627, A63

\bibitem[\protect\citeauthoryear{{\"O}stlin, Hayes, Kunth, Mas-Hesse,
  Leitherer, Petrosian  \& Atek}{{\"O}stlin et~al.}{2009}]{Ostlin09}
{\"O}stlin G.,  Hayes M.,  Kunth D.,  Mas-Hesse J.~M.,  Leitherer C.,
  Petrosian A.,   Atek H.,  2009, The Astronomical Journal, 138, 923

\bibitem[\protect\citeauthoryear{{\"Ostlin, G.}, {Marquart, T.}, {Cumming, R.
  J.}, {Fathi, K.}, {Bergvall, N.}, {Adamo, A.}, {Amram, P.}  \& {Hayes,
  M.}}{{\"Ostlin, G.} et~al.}{2015}]{Ostlin15}
{\"Ostlin, G.} {Marquart, T.} {Cumming, R. J.} {Fathi, K.} {Bergvall, N.}
  {Adamo, A.} {Amram, P.}  {Hayes, M.} 2015, A\&A, 583, A55

\bibitem[\protect\citeauthoryear{{Ouchi}, {Ono}  \& {Shibuya}}{{Ouchi}
  et~al.}{2020}]{Ouchi20}
{Ouchi} M.,  {Ono} Y.,   {Shibuya} T.,  2020, \araa, 58, 617

\bibitem[\protect\citeauthoryear{{Overzier} et~al.,}{{Overzier}
  et~al.}{2008}]{Overzier08}
{Overzier} R.~A.,  et~al., 2008, \apj, 677, 37

\bibitem[\protect\citeauthoryear{Overzier et~al.,}{Overzier
  et~al.}{2009}]{Overzier09}
Overzier R.~A.,  et~al., 2009, The Astrophysical Journal, 706, 203

\bibitem[\protect\citeauthoryear{Pakull, Soria  \& Motch}{Pakull
  et~al.}{2010}]{Pakull10}
Pakull M.~W.,  Soria R.,   Motch C.,  2010, Nature, 466, 209

\bibitem[\protect\citeauthoryear{Pardy, Cannon, {\"O}stlin, Hayes  \&
  Bergvall}{Pardy et~al.}{2016}]{Pardy16}
Pardy S.~A.,  Cannon J.~M.,  {\"O}stlin G.,  Hayes M.,   Bergvall N.,  2016,
  The Astronomical Journal, 152, 178

\bibitem[\protect\citeauthoryear{{Portegies Zwart}, {Baumgardt}, {Hut},
  {Makino}  \& {McMillan}}{{Portegies Zwart} et~al.}{2004}]{PortegiesZwart04}
{Portegies Zwart} S.~F.,  {Baumgardt} H.,  {Hut} P.,  {Makino} J.,   {McMillan}
  S. L.~W.,  2004, \nat, 428, 724

\bibitem[\protect\citeauthoryear{{Prestwich}, {Jackson}, {Kaaret}, {Brorby},
  {Roberts}, {Saar}  \& {Yukita}}{{Prestwich} et~al.}{2015}]{Prestwich15}
{Prestwich} A.~H.,  {Jackson} F.,  {Kaaret} P.,  {Brorby} M.,  {Roberts} T.~P.,
   {Saar} S.~H.,   {Yukita} M.,  2015, \apj, 812, 166

\bibitem[\protect\citeauthoryear{Rangelov, Prestwich  \& Chandar}{Rangelov
  et~al.}{2011}]{Rangelov11}
Rangelov B.,  Prestwich A.~H.,   Chandar R.,  2011, The Astrophysical Journal,
  741, 86

\bibitem[\protect\citeauthoryear{Rangelov, Chandar, Prestwich  \&
  Whitmore}{Rangelov et~al.}{2012}]{Rangelov12}
Rangelov B.,  Chandar R.,  Prestwich A.,   Whitmore B.~C.,  2012, The
  Astrophysical Journal, 758, 99

\bibitem[\protect\citeauthoryear{{Remillard} \& {McClintock}}{{Remillard} \&
  {McClintock}}{2006}]{Remillard06}
{Remillard} R.~A.,  {McClintock} J.~E.,  2006, \araa, 44, 49

\bibitem[\protect\citeauthoryear{{Rivera-Thorsen}, {{\"O}stlin}, {Hayes}  \&
  {Puschnig}}{{Rivera-Thorsen} et~al.}{2017}]{RiveraThorsen17}
{Rivera-Thorsen} T.~E.,  {{\"O}stlin} G.,  {Hayes} M.,   {Puschnig} J.,  2017,
  \apj, 837, 29

\bibitem[\protect\citeauthoryear{{Sazonov} \& {Khabibullin}}{{Sazonov} \&
  {Khabibullin}}{2017}]{Sazonov17}
{Sazonov} S.,  {Khabibullin} I.,  2017, \mnras, 468, 2249

\bibitem[\protect\citeauthoryear{Shapley, Steidel, Pettini, Adelberger  \&
  Erb}{Shapley et~al.}{2006}]{Shapley06}
Shapley A.~E.,  Steidel C.~C.,  Pettini M.,  Adelberger K.~L.,   Erb D.~K.,
  2006, The Astrophysical Journal, 651, 688

\bibitem[\protect\citeauthoryear{Shapley et~al.,}{Shapley
  et~al.}{2015}]{Shapley15}
Shapley A.~E.,  et~al., 2015, The Astrophysical Journal, 801, 88

\bibitem[\protect\citeauthoryear{She, Ho  \& Feng}{She et~al.}{2017}]{She17}
She R.,  Ho L.~C.,   Feng H.,  2017, The Astrophysical Journal, 835, 223

\bibitem[\protect\citeauthoryear{Steidel, Adelberger, Giavalisco, Dickinson  \&
  Pettini}{Steidel et~al.}{1999}]{Steidel99}
Steidel C.~C.,  Adelberger K.~L.,  Giavalisco M.,  Dickinson M.,   Pettini M.,
  1999, The Astrophysical Journal, 519, 1

\bibitem[\protect\citeauthoryear{{Steidel} et~al.,}{{Steidel}
  et~al.}{2014}]{Steidel14}
{Steidel} C.~C.,  et~al., 2014, \apj, 795, 165

\bibitem[\protect\citeauthoryear{Sutton, Roberts, Walton, Gladstone  \&
  Scott}{Sutton et~al.}{2012}]{Sutton12}
Sutton A.~D.,  Roberts T.~P.,  Walton D.~J.,  Gladstone J.~C.,   Scott A.~E.,
  2012, Monthly Notices of the Royal Astronomical Society, 423, 1154

\bibitem[\protect\citeauthoryear{Sutton, Roberts  \& Middleton}{Sutton
  et~al.}{2013}]{Sutton13}
Sutton A.~D.,  Roberts T.~P.,   Middleton M.~J.,  2013, Monthly Notices of the
  Royal Astronomical Society, 435, 1758

\bibitem[\protect\citeauthoryear{{Swartz}, {Soria}, {Tennant}  \&
  {Yukita}}{{Swartz} et~al.}{2011}]{Swartz11}
{Swartz} D.~A.,  {Soria} R.,  {Tennant} A.~F.,   {Yukita} M.,  2011, \apj, 741,
  49

\bibitem[\protect\citeauthoryear{Tenorio-Tagle, Silich, Kunth, Terlevich  \&
  Terlevich}{Tenorio-Tagle et~al.}{1999}]{TenorioTagle99}
Tenorio-Tagle G.,  Silich S.~A.,  Kunth D.,  Terlevich E.,   Terlevich R.,
  1999, Monthly Notices of the Royal Astronomical Society, 309, 332

\bibitem[\protect\citeauthoryear{{Vader}, {Frogel}, {Terndrup}  \&
  {Heisler}}{{Vader} et~al.}{1993}]{Vader93}
{Vader} J.~P.,  {Frogel} J.~A.,  {Terndrup} D.~M.,   {Heisler} C.~A.,  1993,
  \aj, 106, 1743

\bibitem[\protect\citeauthoryear{{Verhamme, A.}, {Dubois, Y.}, {Blaizot, J.},
  {Garel, T.}, {Bacon, R.}, {Devriendt, J.}, {Guiderdoni, B.}  \& {Slyz,
  A.}}{{Verhamme, A.} et~al.}{2012}]{Verhamme12}
{Verhamme, A.} {Dubois, Y.} {Blaizot, J.} {Garel, T.} {Bacon, R.} {Devriendt,
  J.} {Guiderdoni, B.}  {Slyz, A.} 2012, A\&A, 546, A111

\bibitem[\protect\citeauthoryear{{Vito, F.} et~al.,}{{Vito, F.}
  et~al.}{2020}]{Vito20}
{Vito, F.} et~al., 2020, A\&A, 642, A149

\bibitem[\protect\citeauthoryear{Wofford, Leitherer  \& Salzer}{Wofford
  et~al.}{2013}]{Wofford13}
Wofford A.,  Leitherer C.,   Salzer J.,  2013, The Astrophysical Journal, 765,
  118

\bibitem[\protect\citeauthoryear{Yuan \& Narayan}{Yuan \&
  Narayan}{2014}]{Yuan14}
Yuan F.,  Narayan R.,  2014, Annual Review of Astronomy and Astrophysics, 52,
  529

\bibitem[\protect\citeauthoryear{Zezas, Fabbiano, Baldi, Schweizer, King,
  Ponman  \& Rots}{Zezas et~al.}{2006}]{Zezas06}
Zezas A.,  Fabbiano G.,  Baldi A.,  Schweizer F.,  King A.~R.,  Ponman T.~J.,
  Rots A.~H.,  2006, The Astrophysical Journal Supplement Series, 166, 211

\makeatother
\end{thebibliography}




\appendix

\section{Individual Spectral Fits}

\begin{figure*}
	\includegraphics[width=0.80\textwidth]{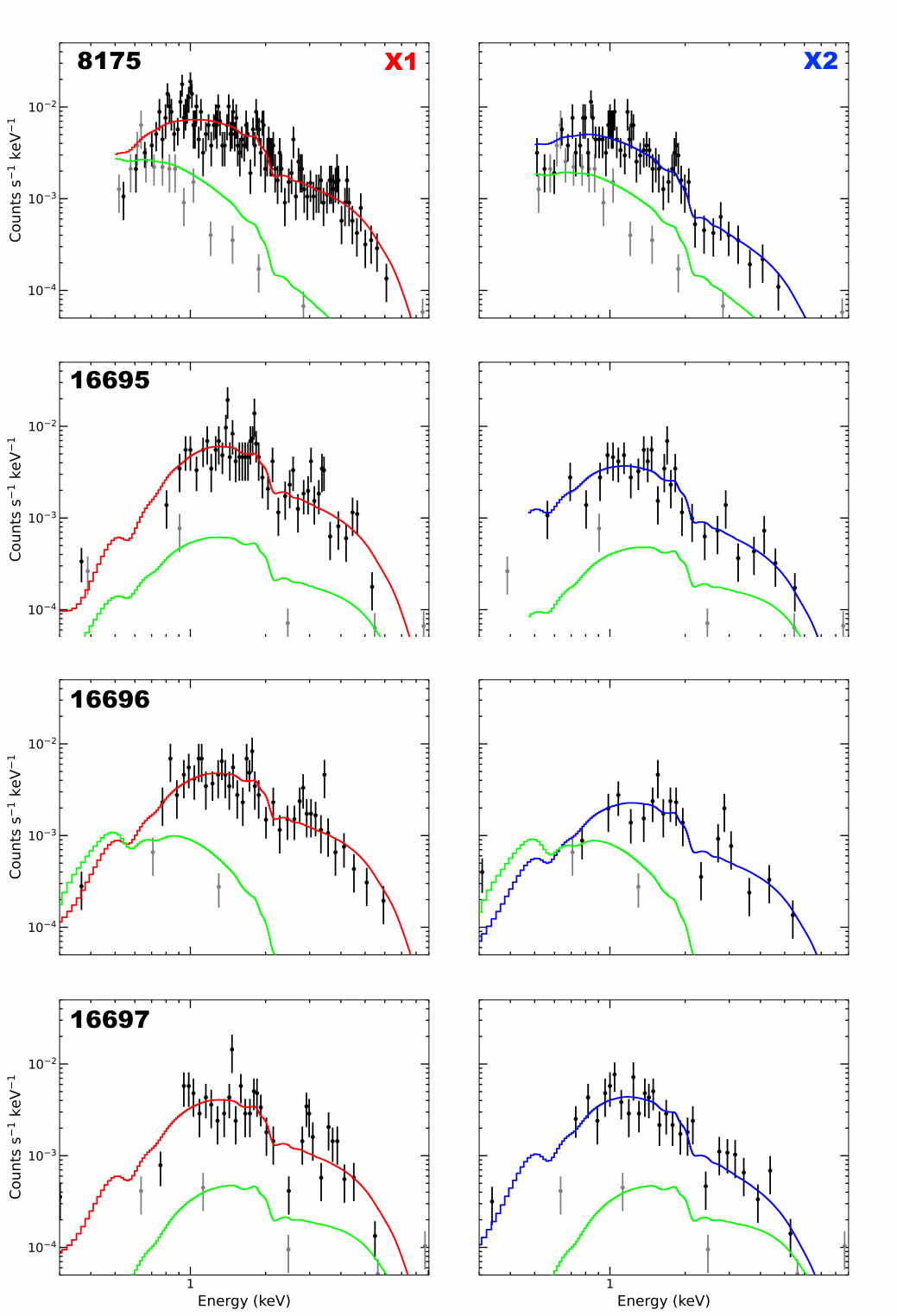}
	\centering
    \caption {Best-fit X-ray spectra of Haro 11 for each individual ObsId. Fits to the region of X1 are shown in the left column, and fits to the region of X2 are shown in the right column.  The data (black points) are binned to a minimum of 5 cts/bin. The background (gray points) is not subtracted, but instead modelled simultaneously (green curve) since Cash statistics are employed. The best fit model (red or blue curve) is an absorbed power law, with best fit parameters and 90\% confidence intervals given in Table \ref{tab:speclums}.}
    \label{fig:spec}
\end{figure*}

\begin{figure*}
	\includegraphics[width=0.80\textwidth]{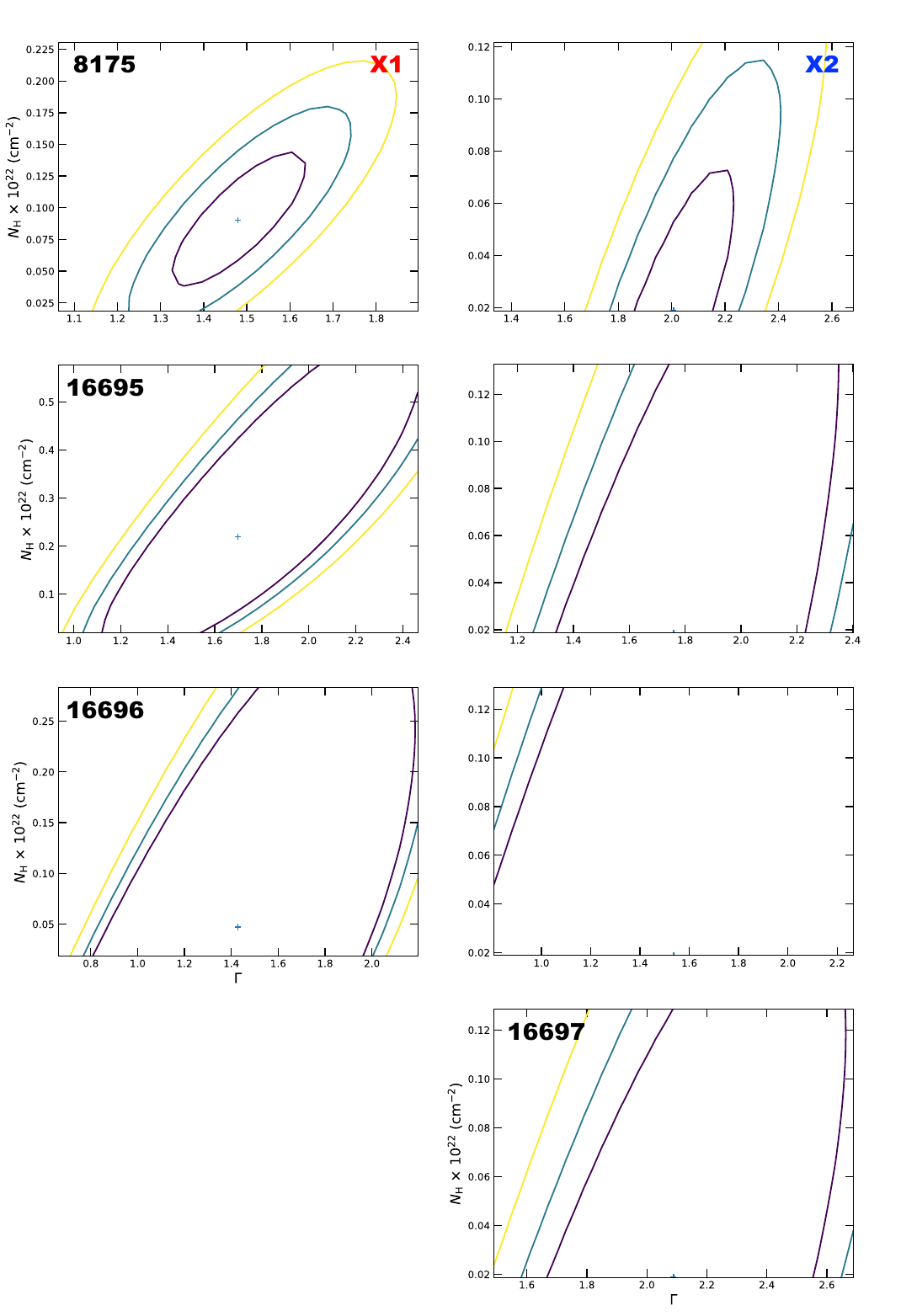}
	\centering
    \caption {Same as the lower panels in Figure \ref{fig:stacked_spec} but for the individual spectra for X1 and X2 shown in Figure \ref{fig:spec}. No error contours are given for X1 in observation 16697 because the parameter {$N_{\rm H}$} was frozen at the galactic value. It is apparent that the spectra with fewer counts have estimated uncertainties that are progressively less well constrained. Even the simple absorbed power law model yields wide uncertainty estimates in most cases. For all fits of X2, the best fit value of $N_{\rm H}$ is at the minimum threshold, set to the galactic value ($N_{\rm H}$ = 1.88 $\times 10^{20}$ cm$^{-2}$). We note that removing the constraint that the absorbing medium cannot be below the galactic value yields even less constrained or unbounded fit parameters for X2, tending towards no absorbing column density. }
    \label{fig:spec_unc}
\end{figure*}



\bsp	
\label{lastpage}
\end{document}